\documentclass[twocolumn,final,natbib]{svjour3ack}

\def\figwidth{\columnwidth}

\usepackage{psfig}
\usepackage{aas_macros}
\usepackage{amsmath}
\usepackage{url}
\usepackage{fixltx2e} %this keeps the figs in numerical order

\newcommand{\plottwo}[2]
           {\centering \leavevmode \psfig{file=#1,width=\columnwidth,clip=}
                            \hfill \psfig{file=#2,width=\columnwidth,clip=}}

\def\lesssim{\lower.5ex\hbox{$\; \buildrel < \over \sim \;$}}
\def\gtrsim{\lower.5ex\hbox{$\; \buildrel > \over \sim \;$}}

\def\msolar{M$_\odot$}

\def\lesssim{\lower.5ex\hbox{$\; \buildrel < \over \sim \;$}}
\def\gtrsim{\lower.5ex\hbox{$\; \buildrel > \over \sim \;$}}

\def\thetamax{75$^\circ$}
\def\batseswiftflu{0.501}
\def\gbmflumult{2.14}

\begin{document}

\title{IACT observations of gamma-ray bursts: prospects for the Cherenkov Telescope Array}

\author{Rudy C. Gilmore \and Aurelien Bouvier \and Valerie Connaughton \and Adam Goldstein \and Nepomuk Otte \and Joel R. Primack \and David A. Williams}

\institute{{\bf Rudy C. Gilmore} 
	\at Santa Cruz Institute for Particle Physics, University of California, Santa Cruz, CA 95064, USA \\ Scuola Internazionale Superiore di Studi Avanzati (SISSA), Via Bonomea 265, 34136, Trieste, Italy
	\\\email{{\em rgilmore@physics.ucsc.edu}}
\and  {\bf Aurelien Bouvier }
	\at Santa Cruz Institute for Particle Physics, University of California, Santa Cruz, CA 95064, USA \\\email{{\em apbouvie@ucsc.edu}}
\and {\bf Valerie Connaughton \and Adam Goldstein }
	\at University of Alabama, Huntsville, AL 35899, USA
\and {\bf Nepomuk Otte }
	\at School of Physics \& Center for Relativistic Astrophysics, Georgia Institute of Technology, Atlanta, GA 30332-0430	
\and {\bf Joel R. Primack \and David A. Williams}
	\at Santa Cruz Institute for Particle Physics, University of California, Santa Cruz, CA 95064, USA}

\authorrunning{R.~C. Gilmore, et al.}
\titlerunning{GRB prospects for CTA}
\date{\today}

\maketitle

\begin{abstract}

Gamma rays at rest frame energies as high as 90 GeV have been reported from gamma-ray bursts (GRBs) by the {\it Fermi} Large Area Telescope (LAT).  There is considerable hope that a confirmed GRB detection will be possible with the upcoming Cherenkov Telescope Array (CTA), which will have a larger effective area and better low-energy sensitivity than current-generation imaging atmospheric Cherenkov telescopes (IACTs).   To estimate the likelihood of such a detection, we have developed a phenomenological model for GRB emission between 1 GeV and 1 TeV that is motivated by the high-energy GRB detections of {\it Fermi}-LAT, and allows us to extrapolate the statistics of GRBs seen by lower energy instruments such as the {\it Swift}-BAT and BATSE on the {\it Compton Gamma-ray Observatory}.   We show a number of statistics for detected GRBs, and describe how the detectability of GRBs with CTA could vary based on a number of parameters, such as the typical observation delay between the burst onset and the start of ground observations.  We also consider the possibility of using GBM on {\it Fermi} as a finder of GRBs for rapid ground follow-up.  While the uncertainty of GBM localization is problematic, the small field-of-view for IACTs can potentially be overcome by scanning over the GBM error region.  Overall, our results indicate that CTA should be able to detect one GRB every 20 to 30 months with our baseline instrument model, assuming consistently rapid pursuit of GRB alerts, and provided that spectral breaks below $\sim 100$ GeV are not a common feature of the bright GRB population.  With a more optimistic instrument model, the detection rate can be as high as 1 to 2 GRBs per year. 

\end{abstract}


%=======================
% 1
\section{Introduction}
\label{sec:intro}
%=======================

The observation of gamma-ray bursts (GRBs) with ground-based atmospheric Cherenkov telescopes has been a tantalizing possibility in recent years.  Powerful $>$10-meter telescope arrays such as H.E.S.S., MAGIC, and VERITAS have come online in the last decade, and satellite detectors such as the {\it Swift} Burst Alert Telescope (BAT) are capable of providing the necessary localization of GRB events within seconds over the Gamma-ray burst Coordinates Network (GCN\footnote{\url{http://gcn.gsfc.nasa.gov/}}).  Despite major campaigns to respond to satellite burst alerts at all three of these instruments \citep{aharonian09,albert07d,garczarczyk08,aune11}, and dozens of follow-up attempts, no conclusive detection of a GRB with an IACT has yet been made.   
Air shower arrays have also played a complementary role in the search for GRBs.  A hint of emission was detected by the Milagrito air-shower array \citep{atkins03}; however no detections were found by the later Milagro experiment \citep{abdo07}.

Prior to the launch of {\it Fermi} on June 11, 2008, knowledge about the emission of GRBs above 100 MeV was limited to a small number of events observed simultaneously in the EGRET and BATSE instruments on the {\em Compton Gamma-ray Observatory (CGRO)} \citep{dingus95,le&dermer09}.  One fascinating 
finding by EGRET was the discovery of an 18 GeV photon associated with GRB 940217, 1.5 hours after the event.  This was a much longer time than the  duration of the burst as measured at lower energies by BATSE, which determined a T90 of 150 seconds\footnote{\url{http://gammaray.nsstc.nasa.gov/batse/grb/catalog/4b/}} (defined as the time between the arrival of 5 percent and 95 percent of the observed fluence).  Though the statistics of these EGRET observations were quite limited, they suggested that high energy emission in GRBs did occur in some fraction of events, and that it could last longer than the lower energy emission.  

{\it Fermi}-LAT, covering the energy range of 20 MeV -- 300 GeV \citep{atwood09}, has now detected emission from over 20 GRBs, of the some 800 detected by GBM at 8 keV -- 40 MeV energy range \citep{meegan09}.  Photons from four of these LAT GRBs were detected above 10 GeV, and 2 above 30 GeV.   At present, the highest energy photons that have been associated with any GRB are a 33.4 GeV photon from long-duration GRB 090902B, and a 31 GeV photon from short-duration GRB 090510.  With the redshift of GRB 090902B determined to be $z=1.822$ \citep{abdo09c}, this implies a rest-frame energy of 94 GeV.   The LAT therefore confirms that emission in the 10 to 100 GeV decade occurs in at least a small fraction of both short- and long-duration GRBs.    However it is not clear how these findings for bright sources extrapolate to the rest of the population, and whether suppression of GeV-scale emission might also happen in a substantial number of cases.

The other major feature of high-energy gamma-ray emission seen by the LAT is the verification of a timescale for the VHE emission that is often longer than that seen by GBM or other experiments sensitive to soft gamma-rays.  An unexpected finding is the delayed onset of emission above 100 MeV, typically by $\sim$10 percent of the GBM T90 duration \citep{dermer10}.  As discussed in \citet{ghisellini10}, emission at high energy is seen to then continue well beyond this time, with a lightcurve described by a powerlaw with slope -1.5.  Understanding the source of this emission, which begins well within the prompt phase of the burst but continues into the afterglow time period, is challenging for models of the high-energy production mechanism.

GeV-scale emission could arise through several mechanisms, and understanding the impact of each on the cumulative spectra will require multiwavelength observations over many orders of magnitude in energy, combined with high event statistics.  High energy emission during the prompt phase of the GRB can be most simply explained by a spectral extension of the internal shock processes (inverse Compton (IC) and synchrotron) that produce the keV--MeV flux \citep{sari&piran97}.  The observed spectrum of GRB 080916C, seen over $\sim 7$ orders of magnitude by the {\it Fermi}-LAT and GBM instruments, could be explained by a constant synchrotron origin \citep{abdo09a}.  Other possibilities include emission of GeV photons from external shocks in the early afterglow of the GRB \citep{fan08} or from the reverse shock formed when the GRB ejecta encounter the interstellar medium \citep{wang05}.  The former can explain the delayed onset of high-energy emission seen in most LAT-detected GRBs.  Several authors have preferred a purely synchrotron origin in the external shock to describe the LAT GRBs \citep{gao09,ghirlanda09,kumar&barniolduran10}, in contrast to one invoking inverse Compton from the shock electrons \citep{zou09}.  Finally, hadronic processes have been proposed as a source of the high energy component, an idea that connects this radiation to the production of ultra-high energy cosmic rays \citep{razzaque09a,asano10}.

The redshifts at which GRBs have been detected span from the local universe to $z=8.2$ \citep{salvaterra09}, corresponding to $\sim$95 percent of the age of the universe.  Confirmed redshifts for LAT-detected GRBs range span a wide range of this distribution from $z=0.736$ (GRB 090328) to 4.35 (GRB 080916C).  This suggests that an IACT-detected GRB could occur at essentially any redshift where star-formation has been observed.  

The cosmological UV-IR background radiation produces a barrier to high-energy photons at extragalactic distances \citep{nikishov62,madau&phinney96}.   Moreover, the large majority of GRBs are believed to exist at high redshift, where the background flux is highly uncertain, and gamma-rays at observed energies as low as 10 GeV can be impacted \citep{gilmoreUV}.  The effective area, angular resolution, background rejection capabilities of IACTs are strongly energy-dependent, and all of these properties decline in quality below a few hundred GeV for current-generation instruments.  Ground-based GRB observations therefore take place in an energy regime where both the low-energy instrument sensitivity and the impact of cosmological background radiation must be carefully taken into account if realistic predictions are to be made.

%TOC
Our goal in this work is not to comment on the preferred emission mechanism for high-energy GRB photons, but rather to build a phenomenological model that best describes this part of the spectrum, based on the limited set of GeV burst detections to date and the much larger body of data available from lower energy experiments.  As the weight of the evidence in the brightest LAT GRBs does indicate the presence of emission mechanisms beyond those producing the prompt flux, we have included allowances for both extended temporal components and separate spectral components from the Band spectrum in this work.  In the following sections we show our predictions for the rate at which CTA will detect gamma-ray bursts, using the information available to us from {\it Fermi}-LAT, lower energy satellite experiments,  and attempted IACT observations with the current generation of instruments.  In \S \ref{sec:grbdet}, we describe the model used in this work.   In \S \ref{sec:results}, we show results for properties of detected GRBs, calculating both the rates at which detections occur in our model and the photon statistics that can be expected from a detection. We use the detection rates by the currently operating {\it Swift}-BAT and {\it Fermi}-GBM instruments as the basis for the calculation.  We also investigate the effect of varying critical input parameters in our model, such as CTA energy threshold, background rate, and telescope response time delay.  In \S \ref{sec:specgrbs} we show the spectra that could be available to CTA from a few sample GRBs with various properties.  The topic of \S \ref{sec:upperlims} is a comparison of our model with the upper limits that have been set by observations to date with IACT instruments, and with the GRB rate observed with {\it Fermi}-LAT.  In \S \ref{sec:disc} we summarize and discuss our findings.

%=======================
% 2
\section{Modeling the Detection of GRBs}
\label{sec:grbdet}
%=======================

To a large extent, the challenge of modeling gamma-ray bursts arises out of the large variance in properties seen between events, and the lack of a simple model describing the radiative mechanism.  In particular, each of the 4 bright GRBs seen by LAT above 10 GeV shows differing behavior. The GeV-scale emission in GRB 080916C, observed by {\it Fermi} three months after launch, was found to be well-described by an extension of the Band function seen at keV and MeV energies \citep{abdo09a}.  Separate spectral components from the Band function were found to be required to match the GeV-scale emission of the three other brightest GRBs in the LAT catalog.  Short-duration GRB 090510 was found to be dominated by a hard spectral component of index -1.62 above $\sim 100$ MeV, compared to the average Band high-energy index of -2.4 \citep{abdo09d,ackermann10}.  Long GRB 090902B was dominated above 100 MeV by emission with a spectral index determined by LAT to be -1.93, compared to an upper Band index of -3.8 \citep{abdo10a}.  The high energy emission also extended in time well past the prompt phase as determined at lower energies, with an only slightly softer spectrum of -2.1 on a timescale of $10^3$~s.  Finally, GRB 090926A, detected up to 19.6 GeV with the LAT, was best fit with a high energy hard component of spectral index of -1.72 and an exponential cutoff at 1.4 GeV, with a high-energy Band index of -2.63 \citep{bregeon11}.

An earlier work, \citet{gilmoreGRB}, addressed the question of detection of GRB photons with the MAGIC telescope and the {\it Fermi}-LAT and the impact of the UV-optical background light.  The basis for this model was the population of GRBs seen by {\it Swift}-BAT, for which redshifts have been confidently determined, and the high-energy statistics of GRBs detected with {\it CGRO}-EGRET.  The flux at the MAGIC energy range was calculated by assuming a power-law spectrum continuing to $\sim 200$ GeV.  Due to the absorption by the extragalactic background light (EBL), photons at higher energy were greatly attenuated and insignificant in number, and the UV-optical EBL therefore has a large impact on GRB detectability.  This calculation showed that MAGIC was capable of detecting tens to thousands of photons from a bright GRB at high redshift, provided that the burst could be observed with a sufficiently low energy threshold.

In this work, we attempt to improve on this previous calculation to make predictions for CTA.  To construct a model for the distribution of GRB properties, we draw upon keV and MeV data from two satellite experiments: {\it CGRO}-BATSE and {\it Swift}-BAT, plus data from the high-energy GRB detections with {\it Fermi}-LAT.  While {\it Swift} has provided us with a large number of GRBs with determined redshifts, the BAT instrument is generally not capable of resolving the Band function peak of the GRB spectra.   We therefore combine the {\it Swift} redshift distribution with the distribution of Band function parameters seen by the BATSE experiment on the {\em CGRO} satellite, at observed energies between 20 keV and 2 MeV\footnote{We preferred using the BATSE catalog instead of the GBM catalog because of its better instrument sensitivity and the much larger number of bursts detected.}.   Band function fits  for BATSE GRBs are taken from the BATSE 5B catalog\footnote{\url{http://gammaray.nsstc.nasa.gov/~goldstein/}} (Goldstein et al. {\it in prep}).   To extrapolate the spectra of these bursts to VHE energies, we use the statistics of GRBs seen by the {\it Fermi}-LAT instrument above 100 MeV  within its first two years of observations.

\subsection{High energy extrapolation}
\label{sec:highengext}
Predicting the GeV-scale emission of GRBs from the well-sampled statistics of lower-energy instruments requires a considerable amount of extrapolation.  Some 4 logarithmic decades in energy lie between the upper extent of the BATSE energy range ($\sim 2$ MeV) and the energy threshold of CTA, which we consider to be between 10 and 25 GeV. We describe our two different approaches to performing this extrapolation in this section.

\subsubsection{Band-function extension model}
\label{sec:bandexdesc}
\begin{figure}
\psfig{file=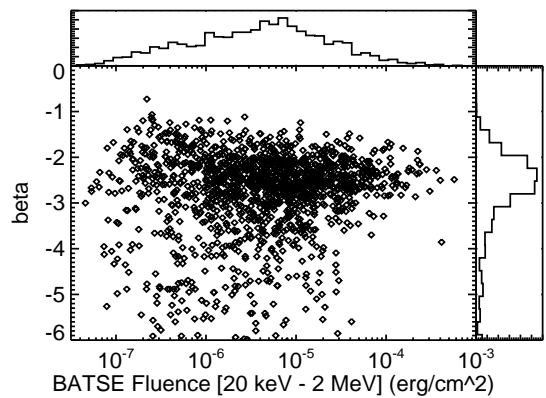,width=\figwidth}
\caption{Band function high-energy indices ($\beta$) vs BATSE fluence for the GRB population, along with distributions for each parameter.}
\label{fig:bandind}
\end{figure}

As a minimal model, we consider the fluence predicted for GRBs at high energy without any significant deviation from the Band fit \citep{band93}:
\begin{align}
\frac{dN}{dE} & \propto  E^\alpha e^{-E(2+\alpha)/E_p}; \;  \; E \leq \frac{\alpha-\beta}{2+\alpha}E_p, \nonumber \\
\frac{dN}{dE} & \propto  E^\beta(\frac{\alpha-\beta}{2+\alpha}E_p)^{\alpha-\beta}e^{\beta-\alpha}; \; \; E > \frac{\alpha-\beta}{2+\alpha}E_p.
\end{align}
\noindent Here $\alpha$ and $\beta$ are low- and high-energy spectral indices, and $E_p$ is the ``peak energy'' describing the location of the turnover.   In this extended Band-function model, termed ``bandex'' in subsequent plots and discussions, the high energy spectrum is assumed to merge seamlessly with the spectral fit determined at lower energy.  A similar model was used in estimating the detectability of GRBs with {\it Fermi} LAT in \citet{band09}.  The high energy normalization is therefore determined by the Band function peak energy and normalization and the upper energy index $\beta$, which continues to GeV energies. In Figure \ref{fig:bandind}, we show the distribution of values for $\beta$ against the BATSE fluence.  A minority of the GRBs in the sample, about 13 percent, have a hard spectrum with $\beta > -2$.  We have enforced the requirement that the total fluence per logarithmic decade not be higher in the GeV range than at BATSE energies, and thus we reset these cases to have $\beta = -2$.  This requirement is consistent with the LAT--GBM fluence relation for long GRBs observed by LAT, but not the short bursts (see Figure \ref{fig:flucomp}), where fluence ratios greater than 1 have been observed.

\subsubsection{Fixed parameter model}
\label{sec:fixeddesc}
\begin{figure}
\psfig{file=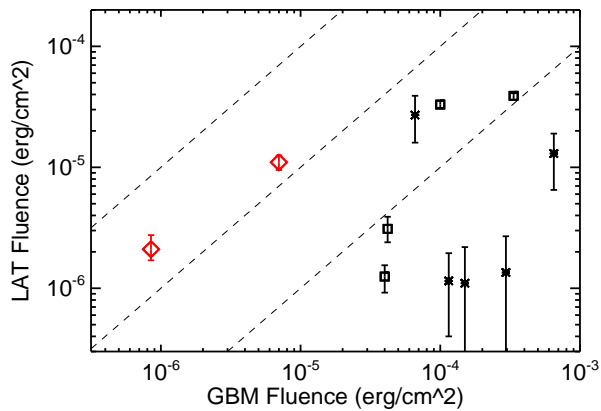,width=8.5cm}
\caption{ A comparison of the LAT and GBM fluences observed in LAT-detected GRBs, reproduced here from \citet{dermer10}.  Black squares are the result for long-duration GRBs, while red diamonds are short GRBs. Dashed lines indicate LAT-GBM fluence ratios of 0.1, 1.0, and 10.0 (bottom to top).  Stars are the analogous EGRET--BATSE fluence relations for bursts detected by EGRET on {\it CGRO}.}
\label{fig:flucomp}
\end{figure}

In this fixed-parameter (``fixed'') model, we make the assumption that the relative fluence between BATSE energies and GeV energies can be described by a single ratio, which we set here to 0.1.  The choice of this parameter is based upon the corresponding ratios found for simultaneously-observed BATSE--EGRET GRBs and for GBM--LAT long-duration GRBs.  In Figure \ref{fig:flucomp}, we reproduce figure 1 in \citet{dermer10}, showing this relation for several {\it Fermi} and {\it CGRO} GRBs.  The spectral index at high energies is set to $-2$, consistent with the mean value for EGRET GRBs of $-1.95$ \citep{dingus95,le&dermer09}, and near the center of the distribution for LAT-detected events \citep{ghisellini10}.  In general, this model requires a significant departure from the extrapolated Band function, and implies the appearance of a separate high-energy spectral component.  As discussed above, such components were seen in GRB 090902B and GRB 090510.  A separate spectral component was also the preferred model in describing the total time-integrated emission from GRB 090926A, albeit with a spectral cutoff of this component at 1.4 GeV \citep{bregeon11}.  Though this cutoff component for GRB 090926A was found with high significance in the integrated fluence, in time-resolved analysis this fit was only preferred over simpler power laws within a single narrow time window.    As GRB 090902B was found to have a fluence ratio of nearly 10 percent and a time-integrated high energy index 1.93 \citep{abdo09c,dermer10},  it can be considered the prototypical GRB in motivating this extrapolation scheme.  While the GeV-BATSE ratio observed for short GRB 090510 was considerably higher, $\sim 1$, we have not included a separate account of the short population because the emission of these GRBs at multi-GeV energies remains very poorly understood, and they are a small part of the {\it Swift} GRB sample, $\sim 9$ percent.  Additionally, the time delay to see GRBs from the ground makes short GRBs very difficult to detect even with a flux factor of unity, and including such an possibility is found to have little effect on our findings.

\subsubsection{High-energy lightcurve}
\label{sec:lci}
As we shall see, the lightcurve and emission duration at GeV energies are critical variables in determining the detectability of GRBs from the ground, where the response time of the telescope to transient alerts from satellite instruments limits observations to $\gtrsim 1$ min after the start of the event, and the background can obscure low-luminosity emission occurring over long timescales.  Motivated by the finding of \citet{ghisellini10}, we will assume that the GRB lightcurve in the early afterglow phase can be described as a power-law falloff.  The prompt phase of the GRB can be demarcated by T90 as determined by lower energy gamma-ray instruments.  Luminosity during this phase is often seen to fluctuate rapidly and unpredictably, with spiked emission features that undergo rapid exponential time decay \citep{piran04}.   Since only considering time-averaged behavior (where the typical erratic prompt emission of GRBs is neglected) will not affect our results in term of detection rate and photon statistics, we describe this phase as having constant flux.  Our total modeled GeV lightcurve then is a plateau from the burst onset ($t=0$) to $t=\mbox{T90}$, followed by a powerlaw falloff:
\begin{align}
\label{eq:lci}
F(t) & =  F_0 ; \; \;  \; \;     & t< \mbox{T90} \\
F(t) & =  F_0 [\frac{t}{\mbox{\scriptsize T90}}]^{-\gamma}; & t \geq \mbox{T90} \nonumber
\end{align}
Here $\gamma$ is the power-law index of the afterglow lightcurve.  We will use 1.5 as a fiducial value, but will also explore the impact of other possibilities in \S \ref{sec:parvar}.  Note that for this value, two-thirds of the total emission emerges after T90.
We assume no spectral evolution between the prompt and afterglow phases.

\subsection{Redshift distribution}
\label{reddist}
\begin{figure}
\psfig{file=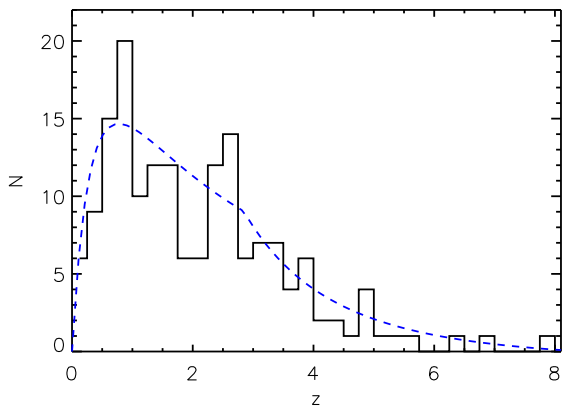,width=8.5cm}
\caption{The redshift distribution determined for {\it Swift} GRBs.  The dashed line shows the fit used in this work.}
\label{fig:swiftzs}
\end{figure}

The observed fluence distribution for GRBs is not assumed to be directly dependent on redshift.  However, redshift is a crucial factor in determining GRB detectability because the cosmological opacity due to EBL is determined by the GRB redshift.  We make the assumption in this calculation that the redshift distribution of GRBs to which CTA responds will be similar to that seen by the {\it Swift}-BAT experiment, which is the only large sample of GRB redshifts available.  Approximately one-third of the GRBs in the {\it Swift} population have well-determined redshifts.  In Figure \ref{fig:swiftzs}, we show the distribution of {\it Swift} redshifts for 167 GRBs, taken from the online Swift GRB Lookup Table\footnote{\url{http://swift.gsfc.nasa.gov/docs/swift/archive/grb_table/}}, along with our fit  to the distribution which is used in most of this analysis.  In \S \ref{sec:gbm}, we will make a speculative alteration to this distribution to describe the redshift distribution of GBM GRBs.

\begin{figure}
\psfig{file=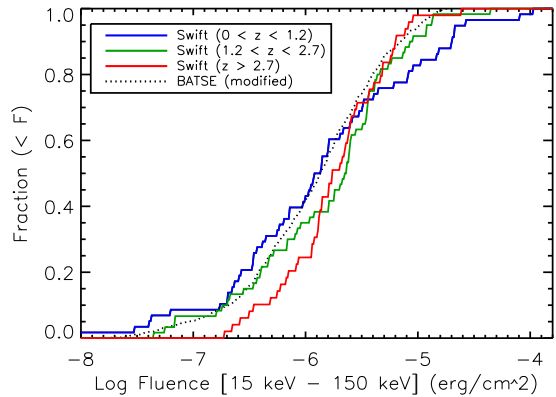,width=\figwidth}
\caption{The cumulative fluence distribution of {\it Swift} GRBs in three redshift bins, which have been chosen to each contain an approximately equal number of events.  Blue line: $z < 1.2$, green line: $1.2 \leq z < 2.7$, red line: $2.7 \leq z$.  The dotted line is the integrated fluence in the total BATSE population over the BAT energy range, 15 to 150 keV, after the correction discussed in \S \ref{sec:batseswiftflumatch}.  Note that this fluence distribution is for all GRBs, not only those with redshift, and therefore it is not necessarily expected to match the {\it Swift} distributions as GRBs with redshift are found to be slightly brighter than the population as a whole (Fig.~\ref{fig:fludistsb}).}
\label{fig:fludistbyz}
\end{figure}

In this work, we make the assumption that redshifts and observed fluence are uncorrelated.  In Figure \ref{fig:fludistbyz}, we show the distribution in fluence for {\it Swift} GRBs divided into tertiles in redshift.  The distribution is not found to evolve strongly in redshift, and the lowest redshift bin actually has the lowest median fluence.   It has been suggested \citep{lloydronning02,salvaterra09a} that luminosity evolution in redshift (e.g., by a factor $(1+z)^{\alpha}$ with $\alpha \gtrsim 1$) is required to best fit the redshift-luminosity relation seen in GRBs.   While the existence and possible origins of such a factor remain controversial, such an evolutionary term could account for our findings in Figure \ref{fig:fludistbyz}.

\subsubsection{{\it CGRO}--{\it Swift} fluence matching}
\label{sec:batseswiftflumatch}
\begin{figure}
\psfig{file=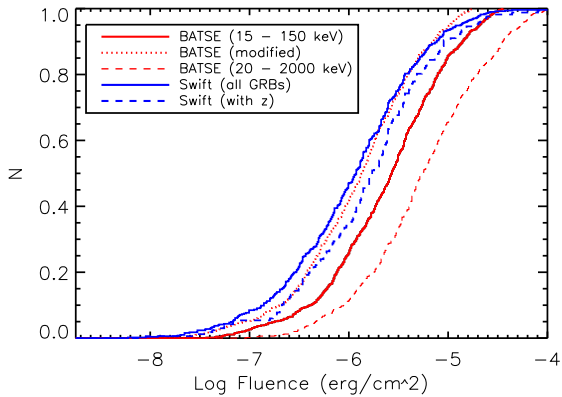,width=\figwidth}
\caption{Integral distibution of {\it Swift}-observed GRBs, compared with the BATSE population integrated over the same energy range.  The red dashed line shows the BATSE population from 20 keV -- 2 MeV, and the solid curve shows these same GRBs integrated over the {\it Swift}-BAT energy coverage of 15 -- 150 keV.  The dotted red curve shows the BAT-band distribution of the BATSE GRBs after the proposed adjustment to best match the distribution of the brightest 50 percent of {\it Swift}-BAT GRBs (solid blue).  Solid and dashed blue lines are the {\it Swift}-BAT fluences for all GRBs and those with redshifts, respectively. }
\label{fig:fludistsb}
\end{figure}

The {\it Swift}-BAT population of GRBs is found to have a lower average fluence distribution than the Band-function fits {\it CGRO}-BATSE 5B population that is sampled to determine GRB fluence and spectral properties for our burst samples, when the latter is integrated over the BAT energy range.  We have adjusted the total fluence of the BATSE burst population by a global factor to better match that of {\it Swift}, which we consider as the GRB trigger instrument in \S \ref{sec:de} and in \S \ref{sec:swiftdetrate}, by using a Kolmogorov--Smirnov test to minimize the difference between the distributions for the brightest 50 percent of bursts in the BATSE sample and the brightest 50 percent in the BAT sample.  The fact that we have restricted our fit to the brightest 50 percent of GRBs is motivated by our finding that GRBs with less than median fluence are not detectable by CTA even under optimal conditions.  The multiplier applied to BATSE fluences in \S \ref{sec:de} is \batseswiftflu.  The distribution of the adjusted BATSE population after this correction is shown alongside the {\it Swift}-BAT fluence distributions in Figure \ref{fig:fludistsb}.

\subsubsection{EBL attenuation}
\label{sec:eblatt}
The EBL, specifically at UV-optical wavelengths, is responsible for attenuating the signal of high energy gamma-rays.  In some EBL models, this attenuation can affect gamma rays at observed energies as low as 20 to 25 GeV for high redshift sources.  The effect of the EBL is to reduce the number of gamma-rays received at high energy, and to reduce the detectability of high-redshift GRBs.  We use as a standard assumption in this work the EBL model and opacities of \citet{gspd11} (GSPD11), based on the semi-analytic modeling of \citet{sgpd11} and \citet{somerville08}.  However, large uncertainties in the EBL normalization are unavoidable at high redshift, and it is useful to see exactly how this uncertainty can influence our predictions.  In \S \ref{sec:eblimpact} we will look at our results assuming a few different models for the background light.

\subsection{Telescope properties}
The second step in constructing our model is a parametrization of the performance of the CTA.  As many of the array properties are indeterminate at the time of writing, we have relied on the design concept for the array described in \citet{ctaconcept10}, as well as reasonable extrapolations from the current generation of IACTs, particularly the MAGIC and VERITAS telescopes.  

\subsubsection{Effective area}
Our assumptions about the effective area of CTA are based on Configuration E, which assumes a central cluster of four 24-meter class large-size telescopes (LSTs) that provide sensitivity to the lowest energy gamma-rays, and an additional 23 medium-size telescopes of the 12-meter class (MSTs) providing sensitivity at higher energies, $\gtrsim 100$ GeV.  Sensitivity at energies above 1 TeV, which is provided by more dispersed arrays of 7-meter class small-size instruments (SSTs), is not crucial to our results here, as most GRBs will occur at redshifts for which emission at these energies is strongly attenuated by the EBL.  

\begin{figure}
\psfig{file=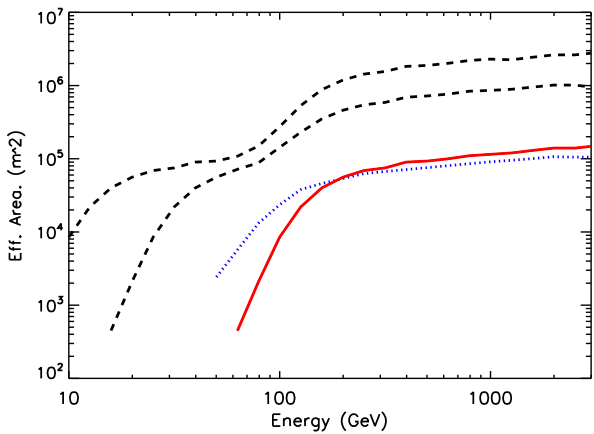,width=\figwidth}
\caption{The effective area functions used in this work.  Solid red is the VERITAS effective area with standard cuts , and the dotted blue line is the MAGIC \citep{albert08a} implementation with standard cuts, shown here for comparison.  The two dotted black curves are the effective area functions used in this work, denoted CTA realistic (lower) and CTA optimistic (upper).}
\label{fig:ea}
\end{figure}

The effective area function of the instrument, after all analysis cuts have been performed, is used to determine the counts per GRB and the significance of the detection presented in the next section.  As the actual function is unknown at this point in time, we assume two functions for the effective area, which are shown in Figure \ref{fig:ea}.   Each of these functions includes contributions from the LST and MST arrays, which dominate the total effective area below and above $\sim 100$ GeV, respectively.  
%We make the assumption that the total effective area at a given energy can be expressed as a direct sum of the contributions from separate arrays. 
For the more conservative of the two, labeled  ``CTA baseline'', the LST contribution is created by shifting the standard VERITAS area function to the 25 GeV threshold that is expected for CTA \citep{teshima11}.  The MSTs use a function that is unmodified in the energy dimension, but has a normalization factor that assumes a linear scaling in effective area with telescope number;  we adopt a factor 25/4 = 6.25 for this case.  We believe these numbers to be a reasonable estimate of the capabilities of the array.  As an alternative, we also present results that incorporate several enhancements to the baseline assumption, and are intended to represent the best possible performance that can reasonably be expected from the instrument.  For this case, labeled ``CTA optimistic'', the LSTs have the same normalization as the CTA baseline array, but with an energy threshold of 10 GeV, which might be achieved through improved trigger and background rejection techniques.  The MSTs are given a normalization 3 times that of the baseline (75/4 = 18.75).   This could either be taken to represent the coverage of a 75 telescope array, or a smaller number of telescopes if the effective area increase with telescope number is found to scale at a faster than linear rate.   The energy shifts for the LSTs in each case are assumed to take place multiplicatively, i.e., $A_{eff}^{CTA}(E)=A_{eff}^{VER}(k*E)$, with $k=4$ for the baseline area function and 10 for the optimistic.

These parametrizations of the telescope effective area refer to a source at zenith.  The changes in telescope performance away from zenith, i.e. at an angle $\theta_z$, are considered using adjustments to the effective area function.  Viewing sources at increasing $\theta_z$ generally introduces a higher energy threshold to the observation, due to the increasing distance between the shower core and the telescope array.  The increased area of the light pool is also the reason why the effective area increases at higher energies.  The first effect is parametrized as a multiplicative energy shift in the effective area function; this is considered as a shift in the telescope energy threshold by a factor $\cos(\theta_z)^{-3}$ away from zenith.  The choice of this index, and the use of a multiplicative shift rather than an additive shift, are both motivated by a fit to the VERITAS effective area at various elevations, with cuts optimized to soft sources (VERITAS Collaboration, private communication).  The second effect is treated by enhancing the effective area at all energies by a multiplier $\cos(\theta_z)^{-2}$, which represents the geometrical increase in the area of the Cherenkov light pool.  The off-zenith effective area is then
\begin{equation}
 A_{eff}^{CTA}(E,\theta_z)=\eta^{-2}A_{eff}^{CTA}(\eta^3 E,\theta_z=0),
\label{eq:zenshift}
 \end{equation}
 \noindent where $\eta=\cos(\theta_z)$ and $A_{eff}^{CTA}(E,\theta_z=0)$ are the at-zenith functions shown in Fig.~\ref{fig:ea}.

\subsubsection{Instrument background}
\label{sec:bkg}
Understanding the instrument backgrounds that will impact GRB observations is critical to predicting the detection rate.  Unfortunately, while published rates from present-day telescopes can be used to predict the background that will impact observations at $\gtrsim 100$ GeV, little is known about how these rates will extend to lower energies.  

We base the background rate in our analysis on that of typical VERITAS observations, using the assumption that the four LSTs will achieve a similar rejection fraction as the four VERITAS instruments above each of their respective energy thresholds, after assuming a power law scaling of the background from 100 GeV to lower energies.   Meanwhile, it is assumed that the MSTs will, at a maximum, have the same background as the VERITAS array multiplied by the scaling in effective area normalization.  The MSTs contribute only a minority of the total background rate in our analyses, so an overestimate of their contribution would have little effect on our results. The background spectrum is assumed to be a power law of index -2.7.  We then take the background at a given energy to scale in proportion to the effective area at that energy.  The total background rate for a telescope set (LST or MST) is then
\begin{equation}
\left(\frac{dN}{dt}\right)_{bkg} = A\int E^{-2.7}\: A_{eff}^{CTA}(E) \; dE    .
\label{eq:bkg}
\end{equation}
The normalizing factor $A$ is chosen so as to produce the  rate for the VERITAS instrument, when this integral is performed over the VERITAS effective area function.  A rate of 6 counts per minute (0.1 Hz) is assumed for VERITAS in the case of the baseline effective area, and this is reduced to 2 counts per minute for the optimistic case.

Using the baseline effective area curve and background rate, together with the scaling described,  we find a differential sensitivity above 100 GeV similar to that shown for configuration E in Figure 24 of \citet{ctaconcept10}.  Below 100 GeV, we find that assuming a $\sim$35 GeV threshold for the LST array in our model produces differential sensitivity similar to configuration E, while the 25 GeV threshold we adopt based \citet{teshima11} gives a sensitivity as much as three times better at some energies.

\subsubsection{Response time}
\label{sec:response}
The transient and random nature of GRB emission represents the main difficulty in detecting emission from these sources.  The onboard satellite localization time of the event, transmission of the data to the ground, the observer's response time, and slew time for the IACT all contribute to a total delay time for the commencement of observation, which we quantify in this work as $T_{delay}$.    The localization time is dependent on the instrument and brightness of the GRB, but times of $<$ 15 sec are typical.   The transmission time of GRB coordinates is expected to be nearly instantaneous \citep{bastieri05}.  The LSTs, which provide coverage at the crucial low energies, are expected to have a slew time of 20 to 30 seconds, while the MSTs may be somewhat slower.   As a standard assumption, we assume a total response time of 60 seconds in this work for the LSTs and 100 seconds for the MSTs, but will also discuss in the next section the effect of varying this parameter.  To date, most observations with the MAGIC telescopes have commenced after considerably longer times despite the instrument's rapid slew capabilities, with only a minority occurring with total delay times of $<$ 100 sec \citep{garczarczyk08}.  It may be that the longer delay times are due to reasons other than the mechanical capabilities of the instrument.  While the inner telescopes of CTA are expected to have generally the same slewing capabilities as those of MAGIC, we will allow for the possibility that future improvements to the GCN and telescope alert procedures and observer response time could lower the typical delay time from current values.

%=======================
% 3
\section{Limits from current experiments}
\label{sec:upperlims}
%=======================

Before discussing our results for simulated CTA observations, it is useful to compare the predictions of our models to the findings of  current-generation GeV-scale experiments.  This enables a test of our model beyond the bright events seen by the {\it Fermi}-LAT.  

\subsection{Analysis of VERITAS GRBs}
\label{sec:verupper}
While upper limits on GRBs have been published for all of the major current-generation IACT experiments, a problem arising when comparing to these limits is the dependence on a particular set of assumptions about the high-energy spectrum, lightcurve, and EBL model.  These must be equivalent for a meaningful comparison to be made between any two sets of observations and/or predictions.  A confirmed GRB redshift is also necessary to determine the impact of EBL on attenuated flux, which can change flux predictions by orders of magnitude.

Here we consider the GRB limits from VERITAS presented in \citet{aune11}.  This work considered 16 GRBs observed by VERITAS over a 30-month period, 9 of which have spectroscopically-confirmed redshifts.  The analysis assumed a characteristic afterglow decay of $t^{-1.5}$ to find the optimal integration timescale for detection, as we have applied throughout this work.  The effect of the EBL was compensated by using the model of \citet{gspd11}, as used in this paper.  Results are presented for gamma-ray spectra of $dN/dE \sim E^{-2.5}$ and $E^{-3.5}$.

To compare our predictions with these results, we have performed a calculation of the high energy emission for the 7 GRBs  from \citet{aune11}  that have both redshift determinations and have been analyzed assuming a $t^{-1.5}$ afterglow decay.  To model these GRBs in such a way that a direct comparison is possible, we use a modified version of our fixed model.  Unfortunately, the flux information provided by {\em Swift}-BAT is generally not sufficient to resolve the spectral peak of the GRB emission and determine the Band function spectral fit.  In our fixed model, we assume that the flux seen in the {\em Swift}-BAT bandpass (15--150 keV) is related to the flux in the 20 keV --2 MeV band by applying the common Band function used in \citet{gilmoreGRB}, which leads to a ratio $\sim 5$ between flux in the 20 keV --2 MeV and {\em Swift} bands.  While this factor is intermediate to the range seen in BATSE GRBs \citep{preece00}, the considerable variations present in the spectral indices and peak energy could change this ratio by a large factor.  To match the spectral index of -2.5 used in the standard analysis, we assume a spectral index of -2, as used in our work, with a spectral turnover to -2.5 at the energy threshold of each GRB observation.

\begin{table*}
\centering
\begin{tabular}{@{}lccccc}

  \hline
  GRB ID & F$_{\mbox{\scriptsize BAT}}$ & Redshift & VERITAS $<$F$>_{\mbox{\scriptsize UL}}$ & $<$F$>_{\mbox{\scriptsize fixed}}$ & $<$F$>_{\mbox{\scriptsize fixed}}$/$<$F$>_{\mbox{\scriptsize UL}}$\\
\hline
070419A & 5.58 & 0.97 &  $2.3 \times 10^{-11}$ & $9.36 \times 10^{-16}$ & $4.07 \times 10^{-5}$ \\
\hline
070521 & 80.10 & 0.553 & $7.6 \times 10^{-12}$ & $1.13 \times 10^{-11}$ & {\bf 1.48} \\  
\hline
080310 & 23 & 2.43 & $ 2.8 \times 10^{-11}$ & $3.27 \times 10^{-17}$ & $1.17 \times 10^{-6}$ \\
\hline
080330 & 3.4 & 1.51 & $3.8 \times 10^{-11}$ & $2.16 \times 10^{-14}$ & $5.69 \times 10^{-4}$ \\
\hline
080604 & 8.0 & 1.416 & $3.1 \times 10^{-11}$ & $8.76 \times 10^{-13}$ & 0.028 \\
\hline
080607 & 240 & 3.036 & $9.8 \times 10^{-11}$ & $5.50 \times 10^{-16}$ & $5.61 \times 10^{-6}$ \\
\hline
090418A & 46 & 1.608 & $6.9 \times 10^{-11}$ & $5.31 \times 10^{-12}$ & 0.077 \\
\hline

\end{tabular}

\medskip
 \caption{Comparison of VHE flux computed for 7 VERITAS GRBs using a modified version of our fixed model (see text) with the upper limits from observation.  Columns show the GRB number, the {\it Swift}-BAT fluence in units of $10^{-7}$ erg cm$^{-2}$ s$^{-1}$, and redshift.  The last 3 columns are VERITAS upper limits, predicted fixed model flux averaged over the observation time, and the ratio of the two.  Cases where predicted flux exceeds the upper limit are shown in bold. } 
\label{tab:verlimits}
\end{table*}

In Table \ref{tab:verlimits}, we compare predictions from our model for GRBs with upper limits placed by VERITAS.  In one instance, GRB 070521, highlighted in the rightmost column, the predicted flux exceeds the upper limit set by VERITAS by a factor of about 1.5.  We do not believe that this case alone poses a problem for our fixed model, as this factor is much smaller than the scatter seen in MeV-GeV fluence ratios for bright GRBs (Fig. \ref{fig:flucomp}), and there is an additional degree of uncertainty in extrapolating the {\it Swift}-BAT bandwidth to the BATSE energy range.  We also note that the $T_{delay}$ and optimal $T_{obs}$ reported in this case are both quite high, 1118 s, and 1809 s, compared to the T90 duration of 38 s.  The predicted high energy emission for this GRB in our model is therefore reliant on the extended-duration lightcurve, up to a timescale of $\sim1$ hour.  Finally, in reporting the redshift for GRB 070521 listed in Table \ref{tab:verlimits}, \citet{hattori07} noted that the detection could be spurious due to a faint afterglow from the supposed host.  If the GRB were at higher redshift, then our predicted flux would be lower and the disagreement lessened or removed entirely.

\subsection{GRB detection with {\it Fermi}-LAT}

{\it Fermi}-LAT provides the largest set of high-energy GRB detections, and we have compared the rate of detections with this instrument with those predicted in our models.  To describe the LAT, we have used the P6 parameters described in \citet{rando09} and on the {\it Fermi}-LAT performance website\footnote{\url{http://www-glast.slac.stanford.edu/software/IS/} \url{glast_lat_performance.htm}}.  Our assumed effective area is based on the ``transient'' event type, and we assume a background rate of 0.05 Hz within the point-spread function of the instrument.  Both the background rate and the effective area at all energies are assumed to evolve as a function of boresight angle uniformly in proportion to the background at 10 GeV.   This analysis uses the same parameters assumed previously, the only change being that there is no delay time in LAT observations, and $T_{obs}$ (the time over which the signal is integrated) is considered on timescales as short as 0.1 s. 

We find detection efficiencies (fraction of GRBs that are detected) above 100 MeV of 12.6 and 5.1 percent for the bandex and fixed models, respectively, for GBM bursts occurring within 70 degrees of the LAT boresight.   The specifics of the calculation used to find detection efficiency will be described in detail in \S \ref{sec:de}.  The fact that these results are inverted from the pattern seen in CTA results in the following sections, where the fixed model generally has a higher detection efficiency, can be explained by the lower energy range covered by the LAT, which favors the softer GRBs in the bandex sample.  These fractions can be compared to the 2-year results of \citet{bissaldi11}, which report 270 GBM GRBs within this angle, and 18 LAT detections; an overall ratio of 6.7 percent.  We have not accounted for autonomous repoints done by the telescope, which have occurred in 45 cases, and could potentially have the effect of raising the detection rate.  GRBs occurring at the center of the LAT field of view are found to have a detection efficiency of about 1.3 times that of all GRBs within 70 degrees of boresight.  As spacecraft repointings have only occurred in a relatively small fraction of cases (45 out of $\sim$540 GRBs), we conclude that the overall impact of these repoints on the detection efficiency is expected to be minor, even before the observational time delay introduced by the telescope slew is taken into consideration.

As discussed in the Introduction, 4 of these detected GRBs have had detected emission above 10 GeV.  In both of our models we find that about 30 percent of LAT-detected GRBs have at least 1 detected photon above 10 GeV, within 1000 sec of the event onset.  This is only slightly higher than the corresponding 2-year ratio of $4/18 \approx 22$ percent.  Due to the long timescale assumed, the model may overestimate the detection rate in some cases due to practical observing constraints, such as instances in which the GRB happens to vanish below the horizon.

%=======================
% 4
\section{Results}
\label{sec:results}
%=======================

\begin{figure*}
\plottwo{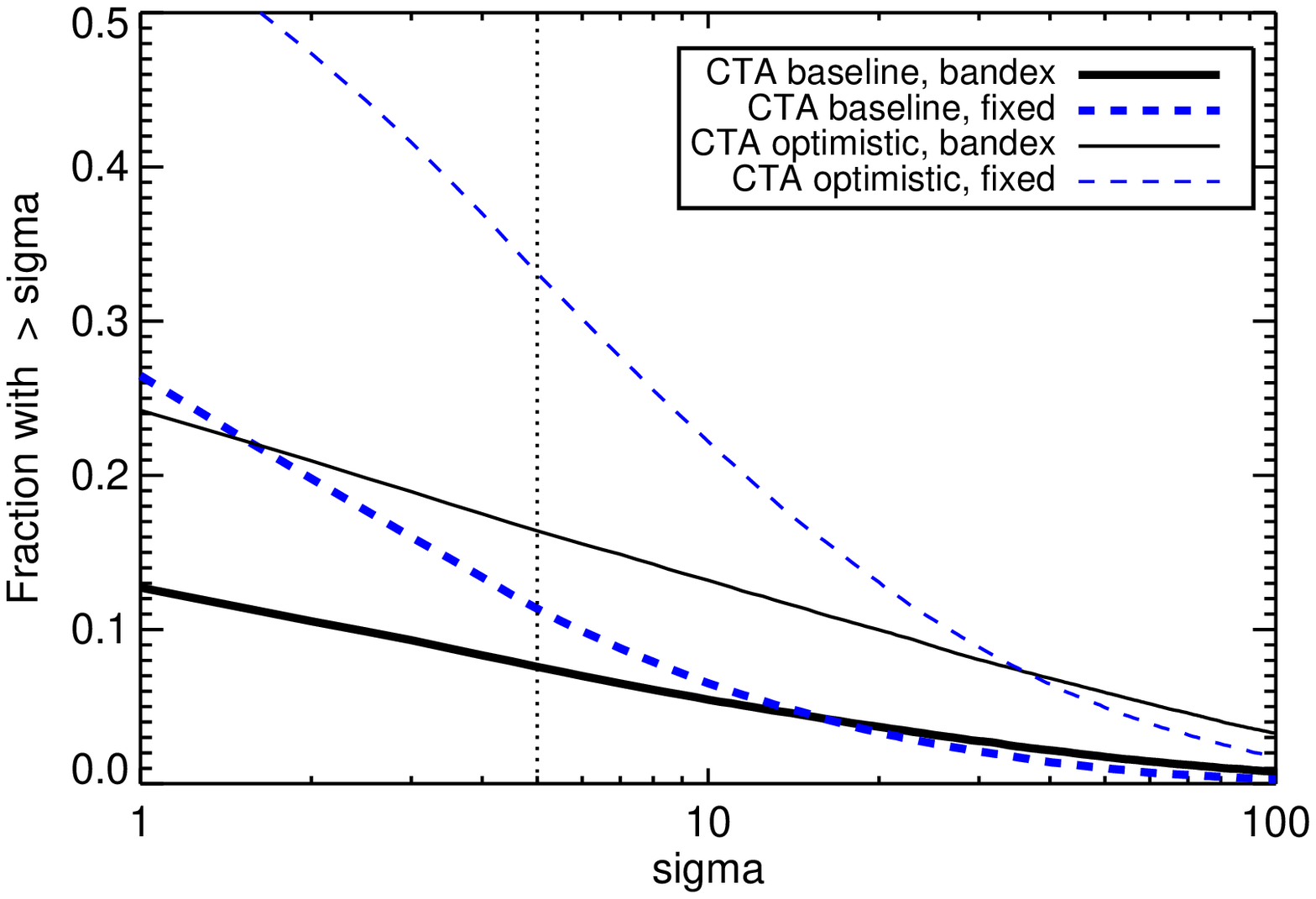}{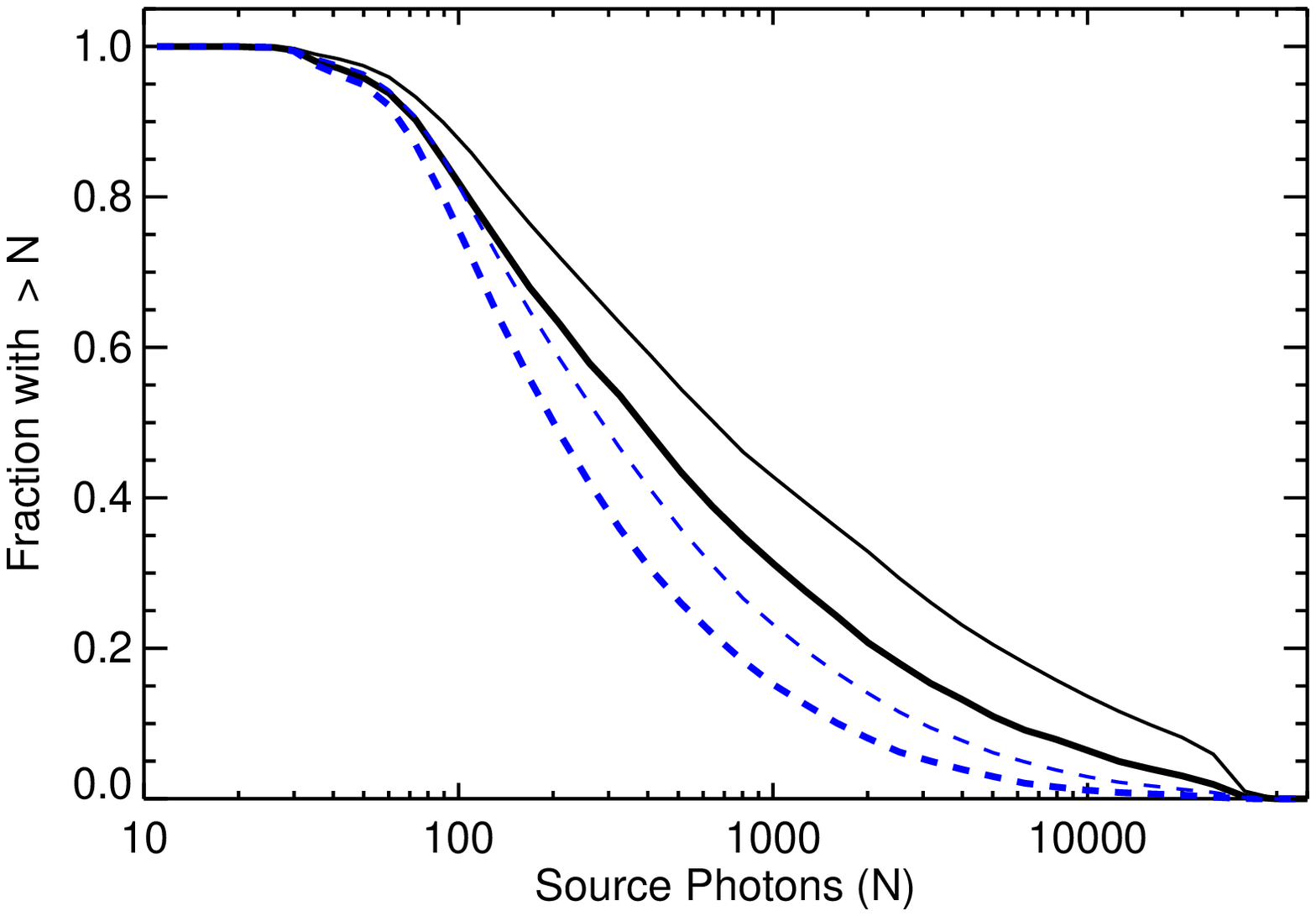}
\caption{Basic statistics of GRBs observed with CTA.  In each panel the solid black line is the result for the direct extrapolation of Band functions (bandex model), from the distribution seen in BATSE GRBs, and the broken blue line is for the fixed model, using parameters described in \S \ref{sec:bandexdesc} and \ref{sec:fixeddesc}, respectively.  Thick lines are for the baseline effective area function, thinner lines are the optimistic function.  {\bf Left:}  The integral distribution of sigma values (significance of the source counts) for all simulated GRBs in the population.   {\bf Right:} The integral distribution of source photon counts for GRBs which are detected, in the timescale bin with maximum sigma.  }
\label{fig:ctasigma}
\end{figure*} 

The GRB detection capabilities of CTA can be described as the product of two independent factors:
\begin{equation} 
\label{eq:dr}
\mbox{Detection Rate} = \mbox{DE} \times \mbox{TR}.  %(\theta_{max})
\end{equation}
\noindent Here DE denotes the detection efficiency, or probability that a randomly-selected GRB for which CTA is able to take data will be detected with a significance of more than 5 standard deviations, and TR is the trigger rate at which the telescope is able to successfully respond to triggers from satellite instruments.  The product of the two is the rate at which confirmed, statistically-significant detections of GRBs will take place.  The factor TR can be decomposed into several independent parameters, which are addressed in \S \ref{sec:detrate}. 

In the following two sections, we show the results of our modeling of the detection efficiency, and its dependence on various instrumental properties.  In \S \ref{sec:de}, we will assume GRBs follow the statistics seen in {\it Swift} detections, and use the redshift distribution and {\it CGRO--Swift} flux multiplier of \batseswiftflu~that was motivated in \S \ref{sec:grbdet}.  In \S \ref{sec:gbm}, we will consider the alerts provided by the GBM instrument on {\it Fermi}, and we will update the redshift distribution and fluence multiplier to better fit this data set.  A particular difficulty arising from GBM alerts is the large positional uncertainty, which is in many cases larger than the CTA field of view, and we devote most of this section to addressing some possible strategies to maximize the usefulness of GBM alerts. 

\subsection{Simulated observations and detection efficiency for {\it Swift}-like GRBs}
\label{sec:de}
For each of the two spectral models considered in \S \ref{sec:highengext}, we consider observations of GRBs randomly placed within a disk of \thetamax~around zenith.  Once the spectrum, lightcurve, and telescope effective area and threshold energy have been determined, we calculate the total integrated counts and background counts over 17 observation timescales ($T_{obs}$), with equal logarithmic spacing from 1 to $10^4$ seconds.  Observations with the telescope are assumed to commence at a time $T_{delay}$ after the beginning of the burst, and end at $T_{delay}+T_{obs}$.   Observed energies are considered from 1 GeV to 1 TeV, although most GRBs experience a spectral cutoff at energies lower than 1 TeV due to the EBL.  For each timescale, we calculate the significance $\sigma$ of the GRB detection, using the  method described in equation (17) of \citet{li&ma83}.  For the purposes of this analysis, an on target -- off target time ratio of $1/3$ is assumed.  The calculated values for $\sigma$ for each timescale are compared, and the highest significance for the bins that have more than 10 photon counts is chosen as the significance for detection of the GRB, and the corresponding $T_{obs}$ is designated the optimal timescale.  The GRB is then assumed to be detected if the significance is more than 5 sigma.

\subsubsection{Distributions in $\sigma$ and $N_\gamma$}

In Figure \ref{fig:ctasigma} we show the basic statistical results of simulated CTA observations for a calculation using the effective area curves of Figure \ref{fig:ea}, $T_{delay}$ = 60 seconds, and a maximum angle from zenith of \thetamax.  A majority of observed GRBs in both models ($\sim$ 90 percent in the bandex model; $\sim$ 80 percent in the fixed) do not lead to a signal of any appreciable significance ($\sigma<1$) for the baseline effective area.  The detection efficiency in the bandex and fixed models is found to be 7.3 percent and 11.4 percent, respectively, for the baseline effective area, and 16 and 33 percent for the optimistic effective area.  Overall, the bandex model shows a flatter distribution of $\sigma$-values than the fixed model.  This is due to the additional degree of freedom introduced in this model by considering the upper Band index in determining the high-energy GRB output in addition to the BATSE fluence, leading to a wider range of values for the overall high-energy normalization.  For the same reason, while the bandex model is more pessimistic in its predictions for detection efficiency, the detected GRBs in this model do often produce more photon counts than the fixed model.  This can be seen in the right-hand panel of Figure \ref{fig:ctasigma}, which shows the distribution of photon counts for detected GRBs.

\begin{figure}
\psfig{file=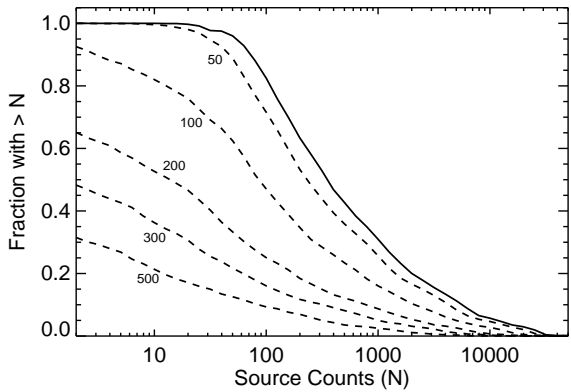,width=\figwidth}
\caption{A breakdown of the number of photons seen in detected GRBs, as a function of energy threshold.  This result is for the bandex model.  The solid curve is the integral distribution for the number of all photons.  From top to bottom, dashed lines shown the distribution of the number of photons above 50, 100, 200, 300, and 500 GeV.}
\label{fig:photcnt}
\end{figure}

Fig.~\ref{fig:photcnt} shows the integral distribution for both total source photon count and integral counts above several energy thresholds for the bandex model with the baseline effective area function.    The majority of photons for most GRBs are seen to arrive below 100 GeV, with a significant fraction below 50 GeV, despite the much larger effective area provided by the MSTs at higher energies.  Also, for a majority of detected GRBs the expected number of source counts above 300 GeV is less than 1.

Table \ref{tab:deinstsum} summarizes the detection efficiencies found for a variety of different possibilities in instrument configuration and assumed GRB population model.  In \S \ref{sec:detrate} we discuss how these results for detection efficiency (fraction of GRBs viewed by the instrument that will be detected) can be converted into a detection rate.  We also show results for the VERITAS array in the table, for comparison.  In this case we follow the same analysis procedure as for CTA, using the VERITAS effective area function shown in Fig.~\ref{fig:ea}, and assuming a delay time of 100 s, rather than 60 s.

\begin{table}
\centering
\begin{tabular}{@{}lcc}
\hline
Instrument &  DE (bandex) & DE (fixed) \\
  \hline
CTA (baseline) & 0.0744 & 0.115 \\
  \hline
CTA (optimistic)  & 0.163 & 0.328  \\
  \hline
CTA (baseline; LST only)  &  0.0732 & 0.110  \\
  \hline
  CTA (baseline; MST only) &  0.0231 &  0.0310 \\
  \hline
 VERITAS ($E_{th} = 65$ GeV) &  0.0241& 0.0281 \\
   \hline
  VERITAS ($E_{th} = 100$ GeV)  & 0.0216 & 0.0235 \\
\hline 
\end{tabular}
\caption{Summary of detection efficiencies for several instrumental arrangements.  In the `LST only' and `MST only', the effective area and background contributions of the MST and LST components are respectively set to zero.  We also show results for the VERITAS effective area, assuming two different energy thresholds.
}
\label{tab:deinstsum}
\end{table}

\begin{figure*}
\plottwo{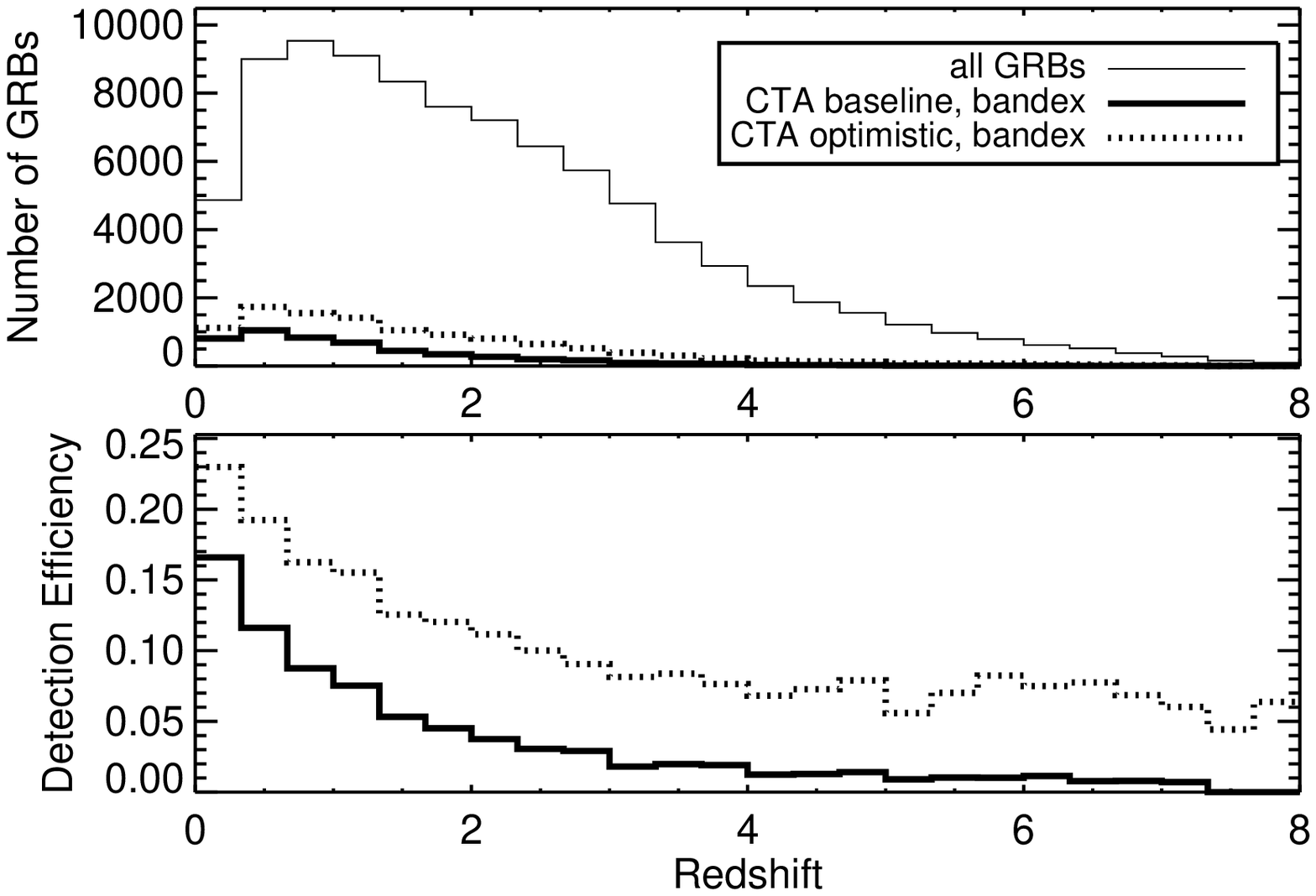}{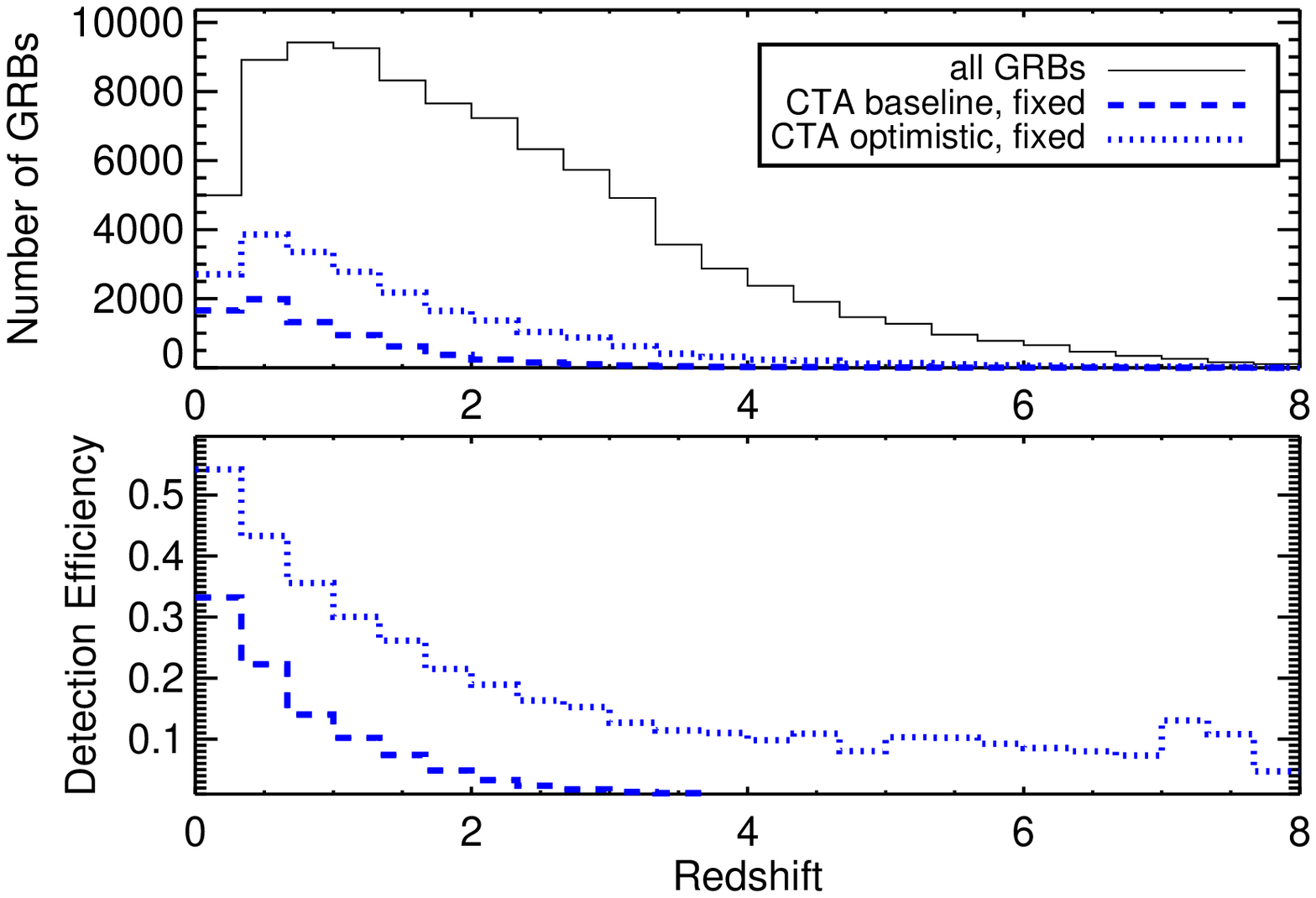}
\caption{The redshift distribution for detected GRBs in our model, for the bandex (on the left) model, and for the fixed model on the right.  The upper panel  on each side shows the number of detected GRBs when a baseline or optimistic (dotted) effective area function is assumed.  The thin black line is the redshift distribution for all GRBs in the sample, which is created from the distribution shown in Figure \ref{fig:swiftzs}.   The lower panel is the fraction of detected GRBs in each bin.}
\label{fig:zdist}
\end{figure*}

Figure \ref{fig:zdist} shows the expected distribution of redshifts for detected GRBs, compared to the whole population.  The CTA effective area, with sensitivity below 50 GeV, potentially allows GRB detections at high redshift, though those at lower redshift will generally have better photon statistics and will therefore be favored.   Assuming the baseline effective area, few GRBs are detected above redshift 2, due to the strong impact at higher redshift of the UV- optical EBL at energies above 50 GeV.  When the optimistic effective area is assumed, a subset of GRBs ($\sim 0.1$) are bright enough from 10 to 50 GeV to be detectable even at very high redshift.  These detections are still a minority of the full set of detected GRBs, however, and are entirely dependent on the low energy performance of the LST array.  In all cases, the distribution is significantly biased towards lower redshifts relative to the {\it Swift} distribution as a whole, with median redshifts of $z=0.9$ and 1.2 for the baseline and optimistic effective area functions, respectively.

In Appendix \ref{sec:otherprops}, we show a number of other properties for GRBs that pass the detection criteria.  This provides some insight into the properties that could be expected of typical IACT GRB detection.  In Appendix \ref{sec:promptonly}, we discuss how our results are affected if only the prompt phase of the burst emission is considered, and the fading afterglow signal is ignored.

\subsubsection{Variation of model parameters}
\label{sec:parvar}
\begin{figure*}
\plottwo{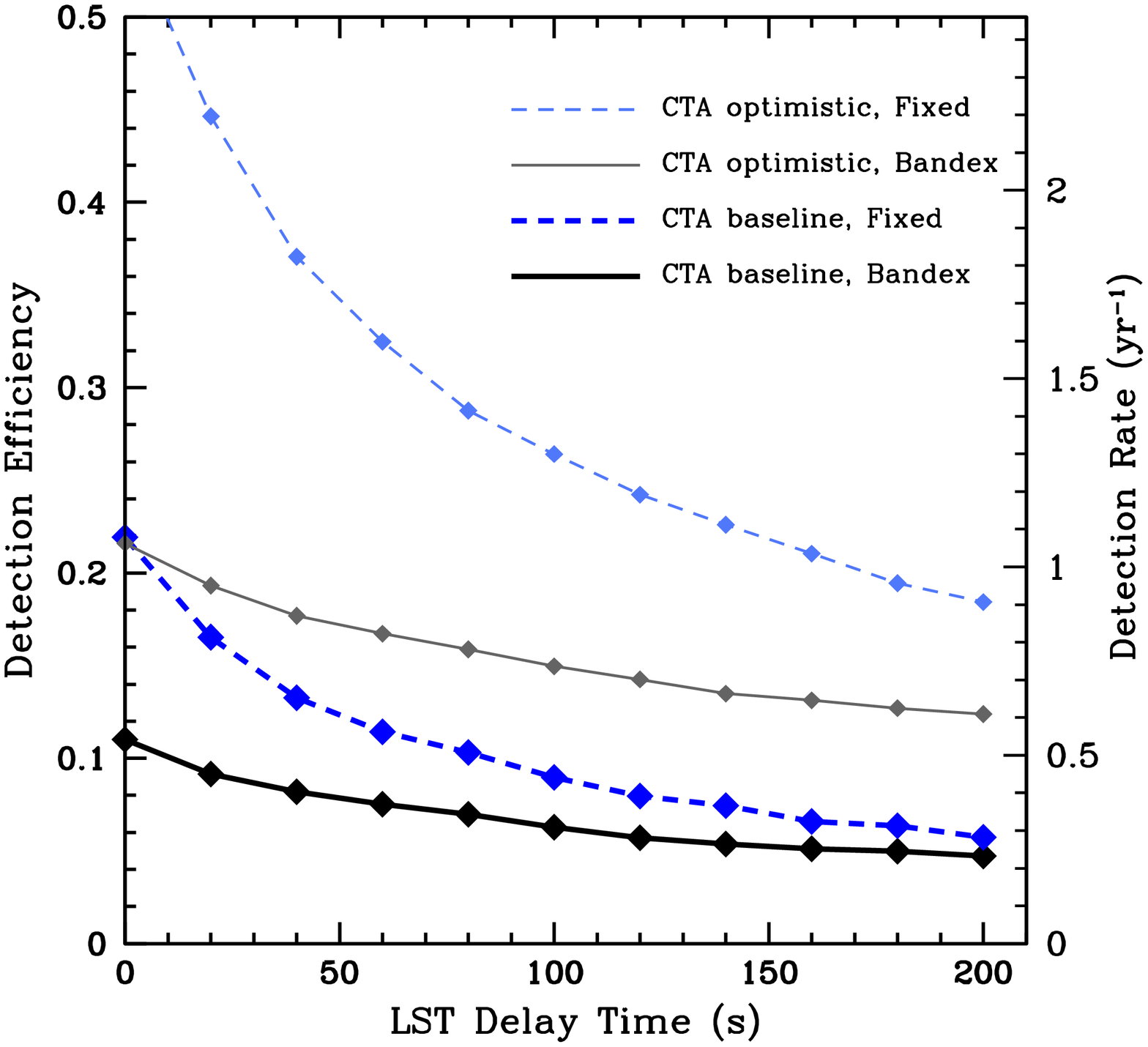}{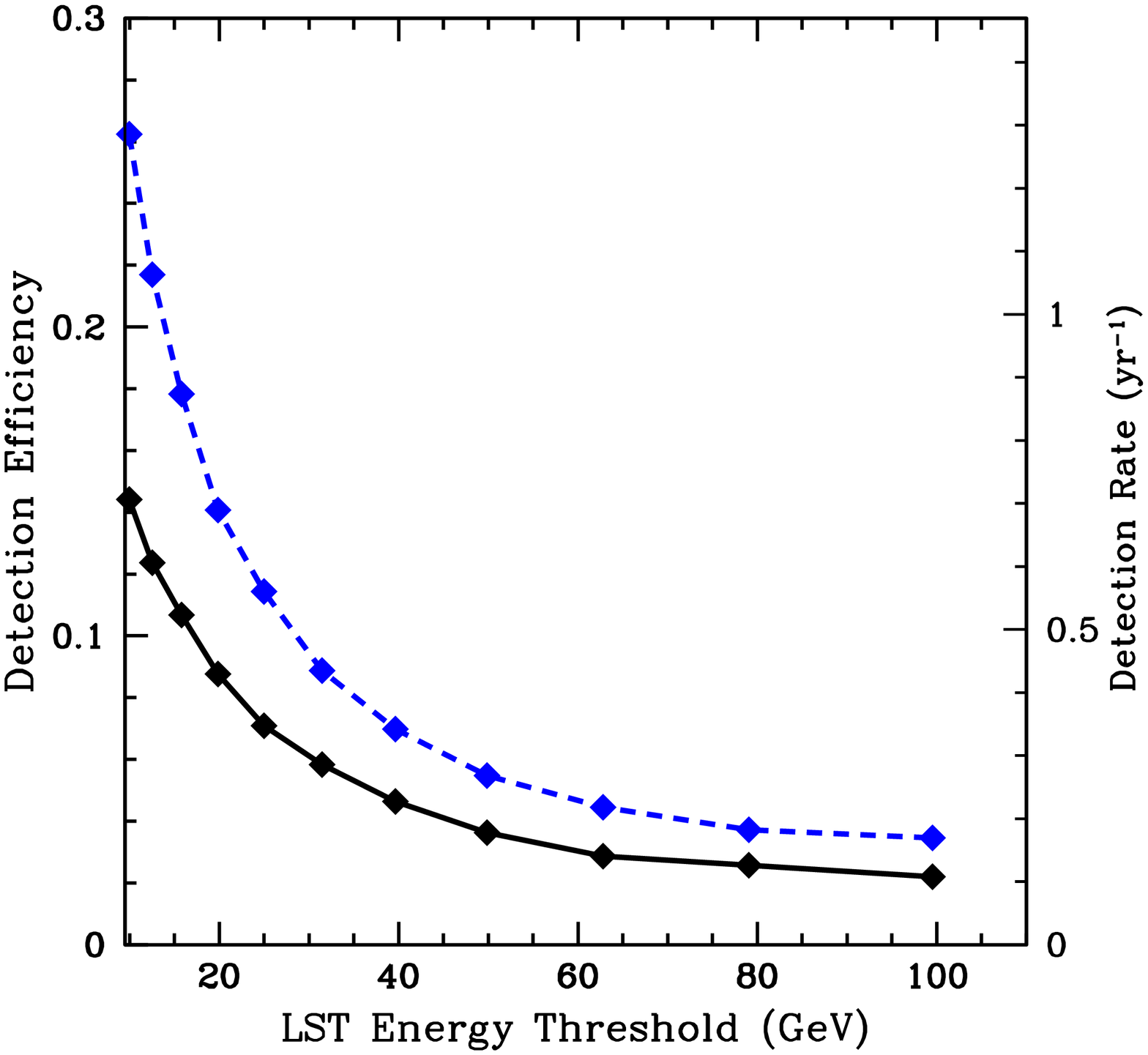}
\plottwo{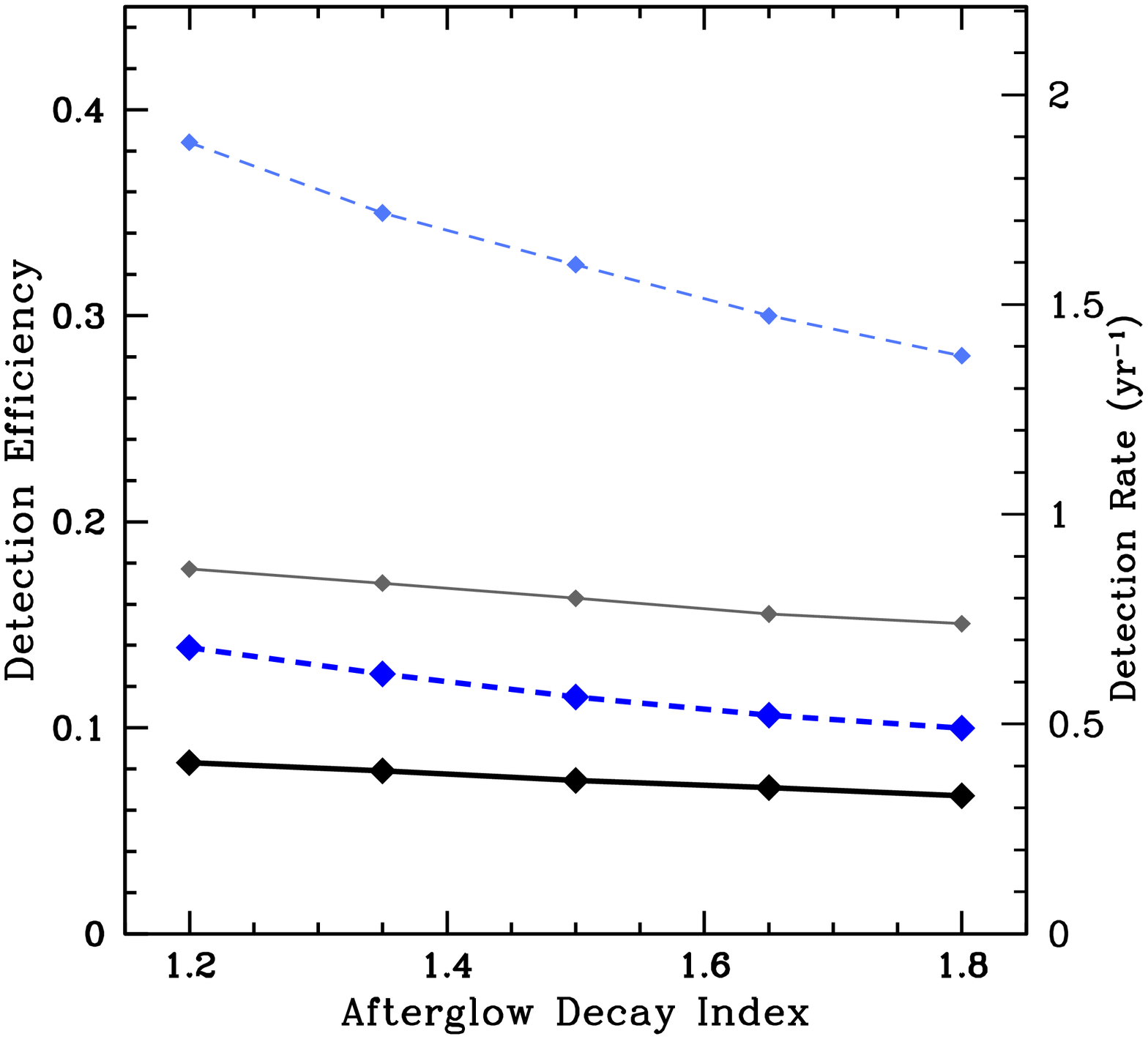}{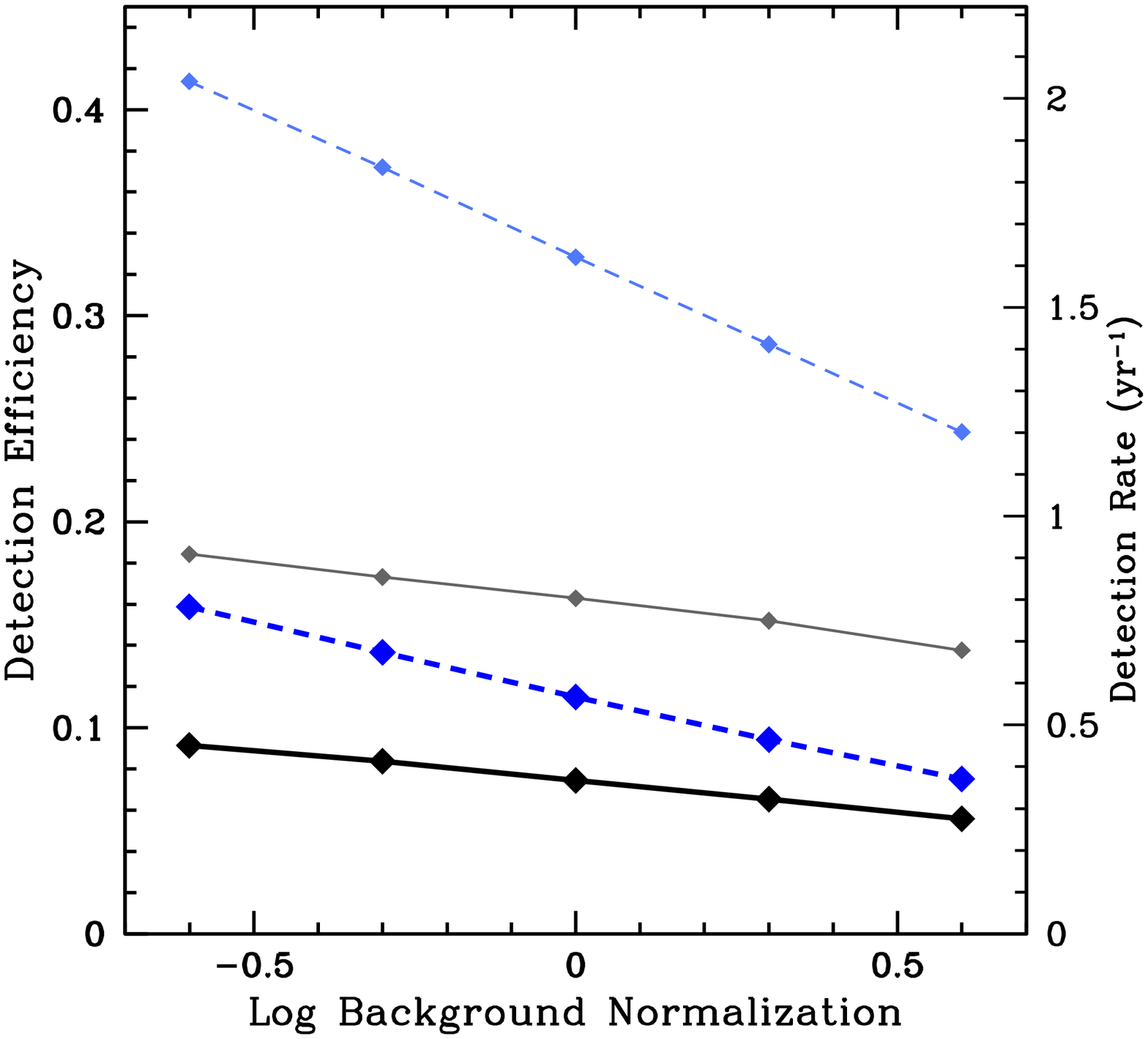}
\caption{Effect on overall detection efficiency and detection rate of varying different assumptions about burst and instrument parameters.  In each plot, the thick solid black and dashed blue lines are predictions from our bandex and fixed model, respectively, for the baseline effective area function.  Thinner grey and cyan lines show the corresponding results using the optimistic effective area.  The detection rates shown on the right-hand axes are based on an assumed instrument duty cycle of 0.1 for all telescopes, and an all sky trigger rate of 95 events/year; see \S \ref{sec:swiftdetrate}.  {\bf Upper Left:}  The effect of varying the observation delay time ($T_{delay}$).  Values are for the LSTs; the MST delay time at each instance is assumed to be the LST time plus 40 seconds.  {\bf Upper Right:} The effect on the overall detection efficiency and rate of changing the energy threshold of the baseline effective area function (Figure \ref{fig:ea}) from its initial value of $\sim 25$ GeV.  The optimistic area function differs from the baseline primarily due to a lower LST energy threshold ($\sim 10$ GeV), and is not shown here.  {\bf Lower Left:} Variations in the assumed lightcurve slope index of the GRB afterglow (Eq. \ref{eq:lci}), and the effect on overall detection efficiency and rate.  {\bf Lower Right:}  The effect on the overall detection efficiency and rate of changing the background rate normalization by a multiplicative factor from its baseline value (parameter `A' in Eq. \ref{eq:bkg}).}
\label{fig:parvar}
\end{figure*}

In this section, we discuss the impact that variations in instrument properties and other general assumptions could have on the GRB detection efficiency.  This demonstrates the effect of variations from our baseline models discussed in the last section.  A summary of results is shown in Fig.~\ref{fig:parvar}.

The impact of VHE observation delay time due to GRB localization and telescope slew time, as discussed in Section \ref{sec:response}, is dependent on the assumed model for the GRB lightcurve at these energies.  The upper left panel of Figure \ref{fig:parvar} shows the overall impact of parameter $T_{delay}$ on the detection efficiency, with other modeled parameters held constant.

Next, in the upper right panel, we show how a higher or lower value of the telescope energy threshold than the $\sim 25$ GeV assumed in the baseline model would influence the detection efficiency.  The effective area function is assumed here to have the same shape as presented in Figure \ref{fig:ea}, but with a shift in energy by a constant multiplicative factor.  As discussed in the introduction, GRB observations are strongly affected by spectral cutoffs due to EBL, and raising the telescope threshold energy reduces the redshift range over which GRBs are detectable.  Detection efficiency is seen here to vary strongly with energy threshold, for both spectral extrapolation models.  Note that setting the energy threshold here to 100 GeV is essentially the same as removing the LSTs from the telescope array (see Table \ref{tab:deinstsum}), since at these energies the effective area function is dominated by the MSTs.  The large decline in detection efficiency with increasing energy threshold demonstrates the importance of having an LST array with low energy threshold to GRB detection, even though the LSTs may only contribute a fraction of the effective area of the total array at higher energy.  

The bottom-left quadrant of Fig. \ref{fig:parvar} addresses how altering the afterglow light curve index $\gamma$ in Eq. \ref{eq:lci} affects results. As discussed in \S \ref{sec:lci}, we have implemented a lightcurve in this work based on the T90 time of a given GRB at Band peak energies, in which VHE emission is flat for this period and then decays as $t^{-3/2}$.  In such a model, $2/3$ of the total VHE energy emerges after the end of the T90 period, leading to a substantial afterglow flux that enables detection of GRBs after the lower energy emission has subsided.  A faster or slower falloff of afterglow flux in time will change the optimal integration time for GRBs in our simulation, as well as the distribution in detection significance and therefore the detection efficiency.  The effect is found to be relatively minor.

Finally, we show in the bottom-right panel of the figure how altering the normalization of the background rate changes detection efficiency.  As discussed in \S \ref{sec:bkg}, the background rate assumed in this work is based on a rate of 6 photons per min over the VERITAS effective area (2 per min for the optimistic effective area), with extrapolation to lower energies achieved with a power law of index -2.7.  This figure shows the effect of variations in this base rate.

\subsubsection{Spectral cutoffs and the impact of the EBL}
\label{sec:eblimpact}
As mentioned in \S \ref{sec:eblatt} and the introduction, the EBL introduces a spectral cutoff in extragalactic gamma-ray observations that affects lower energies at higher redshift.  In Table \ref{tab:eblimpact}, we show how assuming different EBL models can change the overall detection efficiency in our calculation.  In general, the magnitude of UV/optical emission will determine the strength of the spectral cutoffs for GRBs at most redshifts.  However, at $z \gtrsim 2$ this UV emissivity of galaxies is highly uncertain by a factor of several \citep{gilmoreUV} (G09). The first two rows in Table \ref{tab:eblimpact} are the results for the GSPD11 EBL, which we assume elsewhere in this work.  This EBL includes a UV-optical contribution that is nearly maximal in terms of the range allowed by high-redshift measured luminosity functions.  The other rows in the table include models with less UV light; the fiducial model of G09 has a similar star-formation rate to GSPD11, but with more dust attenuation in high-redshift star-forming galaxies that reduces the UV emission.  The low model in G09 has a smaller amount of star formation, in addition to the larger dust extinction.  The model of \citet{franceschini08} (F08) predicts a similar UV flux to GSPD11 at low redshift, but a smaller amount at high redshift.  Running our analysis with the low model of G09 and the F08 model, we find detection rates of GRBs that are about 30 to 40 percent higher than the GSPD11 case.  However, the low model was disfavored in G09 on the basis of IGM ionization data, in favor of the fiducial model, which only increases detection by 5 to 15 percent over GSPD11.  The F08 model only provided gamma-ray opacities for $z \leq 3$, so opacities for high redshift GRBs may be artificially low in this case, and the detection efficiency inflated to some degree.

\begin{table}
\centering
\begin{tabular}{@{}lccc}
\hline
EBL Model & eff.~area & DE (bandex) & DE (fixed) \\
  \hline
 GSPD11 & baseline & 0.0744 & 0.115 \\
  \hline
   GSPD11 & optimistic & 0.163 & 0.328  \\
  \hline
 G09 (low) & baseline & 0.101 & 0.160 \\
  \hline
   G09 (low) & optimistic & 0.189 & 0.425 \\
  \hline
  G09 (fid) & baseline & 0.0803 & 0.130 \\
  \hline
  G09 (fid) & optimistic & 0.171 & 0.365 \\
  \hline
  F08  & baseline & 0.105 & 0.165 \\
\hline 
  F08  & optimistic & 0.192 & 0.423 \\
\hline 
\end{tabular}
\caption{ Detection efficiencies found for a few different EBL models, including the \citet{gspd11} (GSPD11) model used elsewhere in this work, the fiducial and low models of \citet{gilmoreUV} (G09), and the observational model of \citet{franceschini08} (F08).  For the F08 case, gamma-ray optical depths are only reported for $z \leq 3$ and we have used the $z=3$ result for higher redshifts; this result is therefore higher than it might be were the model extrapolated to higher redshift.  The second column shows the effective area function assumed.
}
\label{tab:eblimpact}
\end{table}

\begin{figure}
\psfig{file=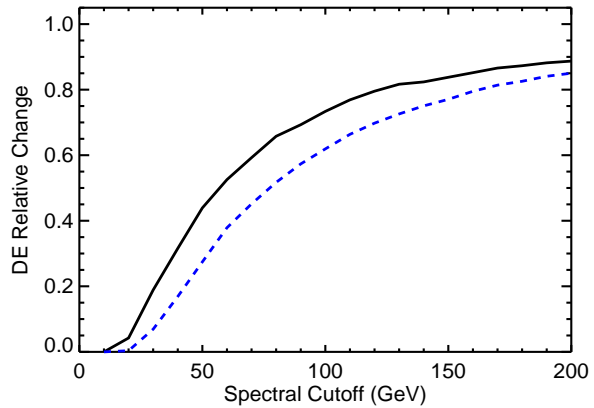,width=\figwidth}
\caption{The reduction in detection efficiency due to the introduction of a universal spectral cutoff at the indicated observed energy.  The vertical axis of the plot is normalized to the detection efficiency found under our standard assumption that emission continues without a break to an energy of 1 TeV.  The solid black line shows results for our bandex model, the dashed blue is for our fixed model.}
\label{fig:scut}
\end{figure}

Another caveat in our analysis is the possibility that GRBs typically have a spectral cutoff or turnover at some characteristic energy.  Due to our lack of knowledge about the GeV-scale properties of GRBs beyond the few bright events that have been detected by {\it Fermi} and {\it CGRO}, it is difficult to explore this possibility in detail.  However, we can perform a simple test, and examine how our results change if a sharp spectral cutoff is assumed to exist at some characteristic observed energy.  Obviously, if this energy is below the sensitivity region of CTA, then the detection efficiency must fall to zero, while if it is above the energy where the EBL has a strong impact on the spectra for the majority of detected GRB, then the effect on results will be minimal.  In Figure \ref{fig:scut}, the impact of a universal step-function cutoff at a given observed energy is shown on the results for the total detection efficiency.

This result suggests that our results will remain sound as long as emission continues unaffected to $\gtrsim 100$ GeV.  In general, GRBs in the bandex model are found to be less strongly affected by this spectral cut, as detected bursts in this scenario include those with softer spectra than the universal spectral of -2 used in the fixed model.  It is worth emphasizing that only GRBs with fluence greater than the median are generally detectable in our simulation (Fig. \ref{fig:ctaflu}), and so the existence of a fluence-dependent cutoff energy that affects only the fainter population of GRBs below 100 GeV would have little effect on our results.

\subsection{Detection of {\it Fermi}-GBM bursts}
\label{sec:gbm}
The GLAST Burst Monitor (GBM, \citealp{meegan09})) on {\it Fermi} presents several unique challenges as a triggering instrument for ground-based follow-up.  GBM is a potent source of GRB alerts, about 250 /yr \citep{paciesas12}, and if the {\it Fermi} mission is extended to a 10-year period, ending in 2018 or later, then there would be significant overlap with CTA operations and many alerts provided under optimal viewing conditions for the array.

Unfortunately, GBM is only able to provide approximate coordinates for the GRB in real time, and the substantial uncertainties are typically similar to or larger than the field-of-view (FoV) of the LSTs, in contrast to the arcminute localizations of the {\it Swift}-BAT.  However, while the analysis of Section \ref{sec:de} assumed a static observation centered on the target, one could also imagine strategies to compensate for the limited field-of-view, at the cost of exposure depth.  After discussing some specifics of our modeled observation of GBM alerts in the following subsection, we will explain one possible strategy for enhancing the DE of GBM bursts.

\subsubsection{Modeling GBM bursts}
\begin{figure}
\psfig{file=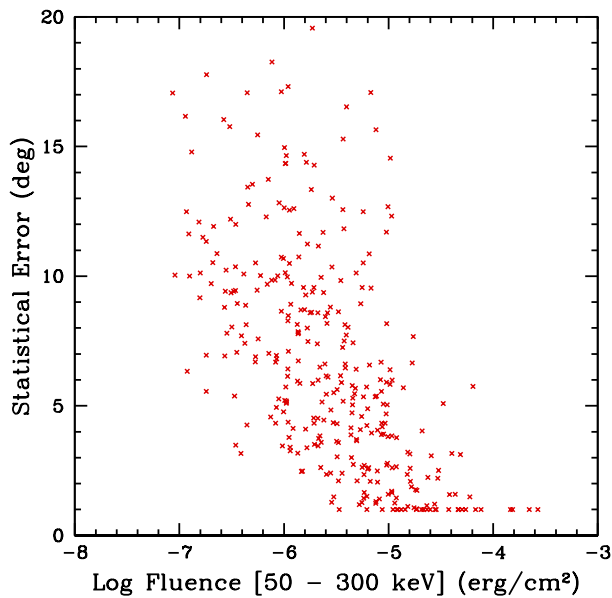,width=\figwidth}
\caption{Comparison of the fluences in the GBM burst population with the individual $1\sigma$ statistical error on the position of each burst.}
\label{fig:gbmbursts}
\end{figure}

Our properties for the population of GBM bursts are taken from a subset of 346 events from the upcoming GBM 2-year catalog (Paciesas et al. {\it in prep}).  Data for these includes the statistical error on the position for each event, and the fluence in the 50 to 300 keV band.  The relationship between the statistical error on the automated localization, produced in ground-processing and distributed within $\sim$10 sec of the GRB trigger time, and the fluence in the 50 to 300 keV band is shown in Figure \ref{fig:gbmbursts}.  The 1 degree lower limit is due to the current grid size of the localization algorithm.

For small statistical errors, a clear negative trend is seen between burst fluence and positional error.  As the brightest GRBs are also the most easily detectable by CTA, this will work to our advantage in the detection efficiency.  GRB positional errors are also subject to considerable systematic uncertainty, which is independent of the statistical error and shows no clear trend in brightness or other GRB properties at this time.  We assume that 70 percent of the GRBs have a systematic error of 3.2 degrees, and 30 percent have a considerably higher value of 9.5 degrees (Connaughton et al. in prep). The total positional error then follows a 2-dimensional gaussian function, with total RMS angular uncertainty given by
\begin{equation}
\sigma_{tot} = \sqrt{\sigma_{sys}^2+\sigma_{stat}^2}.
\end{equation}

The fluence values available for this population cover a different energy range (50 to 300 keV) than the 20 keV to 2 MeV range covered by BATSE.  Our model is based on fluences in the latter energy range.  To compensate, we have calculated the ratios between the fluence over the full BATSE energy range and the 50--300 keV range for the sample of BATSE GRBs.  We then multiply the GBM fluences in the sample by the median of this collection of ratios, to produce a reasonable distribution of fluences over the BATSE energy range for the GBM population.  The ratio found here is \gbmflumult.  

\begin{figure}
\psfig{file=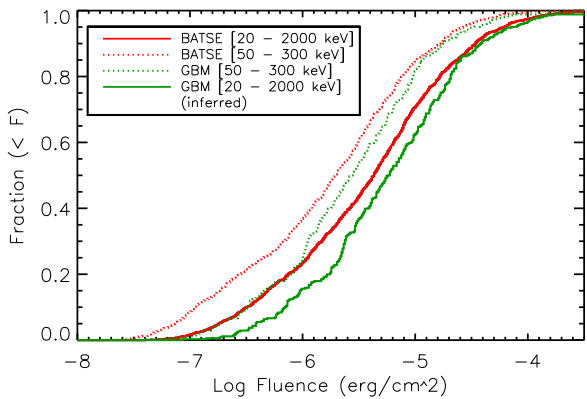,width=\figwidth}
\caption{Integral distributions of fluence values for GRBs observed by BATSE and GBM, including GBM bursts before and after the correction factor described in the text is applied.   Solid red is the distribution of BATSE fluences over the full energy range of the experiment, 20 keV to 2 MeV.  Dotted red are BATSE fluences in the 50 to 300 keV energy band, as determined by integrating over the Band function fits to the bursts.  The green dotted line is the burst distribution in the GBM sample, which is between 50 and 300 keV.   The solid green line then shows the GBM sample after multiplying by the median ratio (\gbmflumult) found for BATSE GRBs between these two energy ranges. Solid lines are therefore a direct comparison of the fluence distributions of the two instruments in our model.}
\label{fig:fludistbg}
\end{figure}

A summary of fluence distributions in BATSE and GBM is shown in Fig. \ref{fig:fludistbg}. Recall that in Section \ref{sec:de}, we applied a multiplier of \batseswiftflu~to account for the differences between the BATSE and {\it Swift}-BAT GRB populations.  Our findings here suggest that GBM bursts are considerably brighter, on average, than those of BATSE, which are in turn brighter than {\it Swift} detections (even the sub-set of {\it Swift} GRBs with known redshifts).  

\begin{figure}
\psfig{file=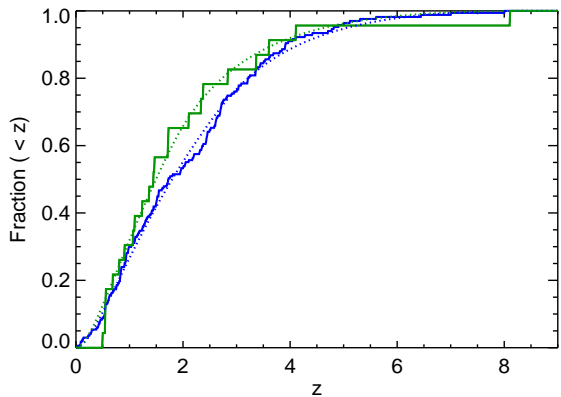,width=\figwidth}
\caption{The integral distribution of redshifts in all {\it Swift} GRBs (blue) compared with the subset that also have GBM detections (green).  Solid lines are the data for each, while the dotted blue and green lines are the fits used in the previous sections and the following sections, respectively.}
\label{fig:swiftgbmzcomp}
\end{figure}

As GBM has substantially different energy coverage from {\it Swift}-BAT, and detects GRBs that are considerably brighter, one might expect the redshift distribution of GBM-detected GRBs to differ from the {\it Swift}-BAT population.  Unfortunately, there are not enough known redshifts within the GBM population to do a comprehensive analysis of the differences.  This is largely due to the positional errors on GBM detections which make follow-up observations impossible without more accurate data from another experiment.  We can, however, look at the handful of GBM bursts that are also listed with redshifts in the {\it Swift} catalogue.  A plot of redshifts for these 23 GRBs is shown compared to the redshift distribution of all {\it Swift} GRBs in Fig.~\ref{fig:swiftgbmzcomp}.  GBM bursts are found to have a somewhat lower distribution of redshifts overall, and we have made a modification (dotted green line in the plot) to the fit for the {\it Swift} distribution that we will use in the analysis that follows.

A final factor we must consider is an account of the camera sensitivity lost in off-axis CTA observations.  We account for this by including a simple sensitivity factor that is a function of radius ($r$) from the FoV center:
\begin{align}
S(r) & =  1;  r < R_{fov}-1.5 \nonumber \\
S(r) & =  0.3(R_{fov}-0.5-r)+0.7; \nonumber \\  & \hspace{2cm} R_{fov}-1.5 < r \leq R_{fov}-0.5 \nonumber \\
S(r) & =  0.4(R_{fov}-r)+0.5; R_{fov}-0.5 < r \leq R_{fov} 
\end{align}
\noindent Here $R_{fov}$ denotes the maximum extent at which any observation is possible, and is equal to half the FoV value.  This radial dependence gives full sensitivity up to $R_{fov}-1.5$, 70 percent sensitivity at $R_{fov}-0.5$ and 50 percent sensitivity at the edge of the camera.  This factor is applied to both expected number counts from the GRB and the background rate, and is used for both LST and MST observations.   We set $R_{fov}$ to 4.25 degrees for the LSTs and 8 degrees for the MSTs.

\subsubsection{Static observations of GBM bursts}
\label{sec:statmode}

As a first step, we calculate the detection efficiency of this population without applying any search mechanism in the GRB observation.  This analysis, and that of the next section, will only utilize the fixed model, as spectral details for this burst population are not available at the time of writing.  Results from the previous sections indicate that detection efficiencies of GRBs are typically a factor of 1.5-2 times higher in the fixed model than in the bandex model; there is no reason to believe a similar relationship would not hold true as well here.

\begin{figure}
\psfig{file=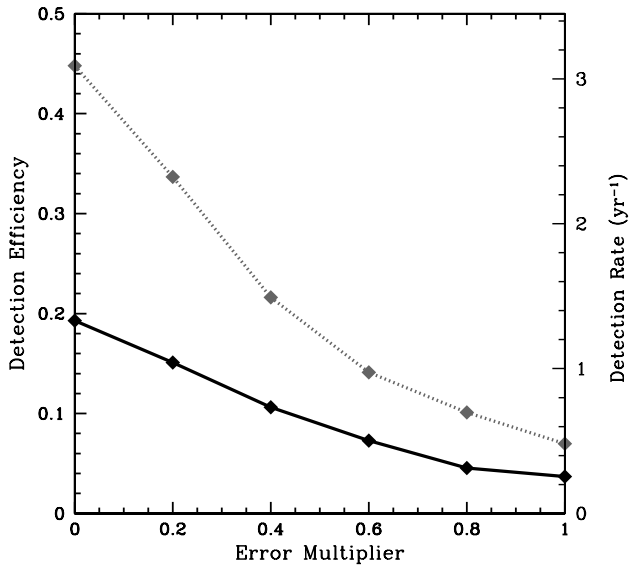,width=\figwidth}
\caption{Detection efficiency for static observations of GBM bursts, as a function of the uncertainty in burst position.  Variations along the x-axis indicate the effect of multiplying the total positional uncertainty for each GRB, which is calculated as described in the previous section, by a given constant scaling factor.  Curves show the detection efficiency for the CTA baseline (solid black) and CTA optimistic (dotted grey) effective area functions.  The right-hand axis shows the detection rate for the telescope under a standard set of assumptions; see \S \ref{sec:detrate} for details.}
\label{fig:gbmerrred}
\end{figure}

The limiting factor in these observations is the field of view of the telescope.  As most GRBs in the sample have a minimum uncertainty of $\sim 4$ degrees from combined statistical and systematic effects, a 3 to 5 degree diameter FoV is insufficient to catch more than a minority of events.  Figure \ref{fig:gbmerrred} shows the current detection efficiencies calculated, and how these could be increased by future reductions in the amount of uncertainty affecting GBM burst positions as reported in real time.  

\subsubsection{Scanning mode observations}
\label{sec:scanmode}
 A possible solution to the FoV limitations of CTA is to attempt to rapidly scan over some portion of the GRB error box after the burst alert, rather than simply observing the coordinates of the best GRB localization.  We consider in this section the possibility of increasing the GRB detection efficiency using ``scanning mode'' observations, and address the question of how the search box should be chosen to optimize this rate.  A larger search box increases the probability that the GRB will be observed, but at the cost of exposure time.  We make the assumption here that the data from such a scanning observation over minutes or hours could be compared with a later determination of the actual burst position, and the significance of the GRB detection would then be computed in an after-the-fact analysis taking into account the photon counts and expected background within the source PSF.  We do not consider here the possibility that VHE emission could be identified in real time, i.e., for a ``stop-on-target'' type of scan.

Our calculation of the detection efficiency follows that of \S \ref{sec:de} above, with a few modifications.  We assume that the time required for the LST array to search a region of 15 degrees radius is about 120 seconds.  We therefore append the delay time used in the last section with another term that describes the delay  between the commencement of the scan and when the telescope first passes over the GRB's true location.  The total delay is then
\begin{equation}
T_{delay}=60 + R_N(0,120) \frac{(R_{srch}-R_{fov})^2}{(15^\circ-R_{fov})^2} \: \; \mbox{sec},
\end{equation}
where $R_{srch}$ is the radial extent of the search box, $R_{fov}$ is one half the field-of-view for the LSTs, and $R_N(0,120)$ is a random variable between 0 and 120.  Afterwards, it is assumed that the telescopes make many successive passes over the GRB, and that the integrated exposure after $T_{delay}$ is approximately equal to that of the standard calculation multiplied by a factor $(R_{fov}/R_{srch})^2$ if the GRB is in the search box, zero otherwise.  The received background is adjusted by the same factor.  The MST array is assumed to search in the same pattern as the LSTs, and their total exposure is multiplied by an analogous factor.

We consider several possible ways that one might determine the extent the scanned region, $R_{srch}$.  The simplest possibility is to use a constant for all GRBs.   In a scan over possible values in steps of 0.5 degrees, we find that detection efficiency is maximized at $R_{srch}$ = 6 degrees for the baseline effective area, and 7.5 degrees for the optimistic function.  For the baseline effective area, the detection efficiency is increased to 0.057 for $R_{srch} = 6$ degrees, compared to 0.037 in the static case; an increase of over 50 percent.  In the optimistic case, the detection efficiency can be more than doubled, from the static value of 0.07 to 0.16 with an optimized scan box.  These results show that a scanning strategy for GRB follow-up can raise the detection probability by a large factor, and that it is especially powerful for a telescope with a very low energy threshold.

\subsection{Total detection rate}
\label{sec:detrate}

The second part of our calculation of the detection rate, summarized in Eq.~\ref{eq:dr}, is an estimation of the trigger rate from satellite instruments.  This factor accounts for the sky coverage and duty cycle of the instrument, and is the rate at which CTA can respond to and observe GRBs.  All other factors influencing GRB detectability are incorporated into the detection efficiency parameter discussed above.   Our calculations in the last section allow for observations at a maximum angle from zenith of 75 degrees, covering 37 percent of the sky.  The calculations of detection efficiency include the effect of increasing energy threshold for observations far from zenith.   The duty cycle of Cherenkov telescopes is limited by the requirement that these telescopes operate mostly on clear, moonless nights, which has generally produced realistic values of about 10 percent, and we expect this factor to remain valid for CTA.  Operation of CTA during moonlight may increase the duty cycle to $\sim$13 percent or more, but at the cost of a higher energy threshold.  We therefore do not expect a significant change of our predictions for the overall detection rate.

The total trigger rate calculated from a given satellite alert rate 
\begin{equation}
TR = DC \times SC \times SR \times BF.
\label{eq:tr}
\end{equation}
Where the CTA duty cycle $DC$ is $\sim 0.1$ and the sky coverage factor $SC$ is 0.37, for a 75 degree radius area around zenith, as we have assumed throughout this work.  The satellite rate $SR$ is the number of GRB alerts produced by a given satellite detector per unit time.  The bias factor ($BF$) includes the effect of any correlation or anti-correlation between the location of GRB satellite alerts and the sky area covered by CTA.  This includes factors such as the bias present in {\it Swift} GRB alerts \citep{gilmoreGRB}, which leads to GRB discovery preferentially in the anti-solar direction, working to the advantage of IACTs which are limited to nighttime operations.  A factor of 1 indicates no departure from a random distribution of GRB alerts on the sky.  We assume a factor of 1.4 for {\it Swift}-BAT GRBs and 1.0 for GBM. 

\subsubsection{{\it Swift}-like GRBs}
\label{sec:swiftdetrate}

The {\it Swift} satellite, launched in late 2004, has detected GRBs at a rate of about 95/yr over its first 70 months of operation, and is expected to have an orbital life of $>$15 years \citep{romano10}.  If the science lifetime of {\it Swift} overlaps with that of CTA, then {\it Swift} will provide a constant source of well-localized GRB alerts.  Using Eq.~\ref{eq:tr}, we estimate a detection rate for {\it Swift} alerts of \[DR_{\mbox{\tiny Swift}} = DE \times 4.92 \;\frac{\mbox{\small GRB}}{\mbox{\small  yr}}.\]  If we assume the `best-guess' instrument parameter of $T_{delay}=60$ sec for the LSTs and use the detection efficiencies from the first two rows of Table \ref{tab:deinstsum}, we find 
detection rates of 0.37 and 0.57 yr$^{-1}$ for the bandex and fixed models with the baseline effective area function, and 0.80 and 1.61 with the optimistic effective area function.  These correspond to timescales of 32 and 21 months between GRB detections with the baseline area, and 15 and 7.5 months with the optimistic effective area functions.

Another upcoming mission that could provide timely GRB localizations is the {\it SVOM} satellite.  As described in \citet{gotz09}, {\it SVOM} will consist of an orbiting gamma-ray telescope covering an energy range similar to that of {\it Swift} that is expected to detect $\sim$70 GRBs/yr, as well as ground-based telescopes for follow-up observations.  Because the ECLAIRs/CXG instrument on {\it SVOM} is intended to cover a similar energy range to {\it Swift}-BAT, the population of GRBs detected with this satellite should be similar to the {\it Swift} population, and therefore have a similar detection efficiency to that which we have calculated here.   

\subsubsection{{\it Fermi}-like GRBs}

GBM on {\it Fermi} has detected GRBs at a rate of about 250 per year (Paciesas et al. {\it in prep}).  
Because these GRBs are detected at all points above the horizon, the anti-solar bias factor affecting {\it Swift} (and presumably {\it SVOM}) does not apply here.  We can therefore write the detection rate for GBM GRBs as \[DR_{\mbox{\tiny GBM}} = DE \times 9.25 \;\frac{\mbox{\small GRB}}{\mbox{\small  yr}},\] meaning that GBM should provide around 10 alerts per year that can be investigated by CTA.  While these GRBs are brighter on average than {\it Swift} GRBs, ground-based followup is hampered by the large uncertainty in burst location, which is generally several degrees.  In \S \ref{sec:scanmode} we showed that the detection efficiency of GBM alerts can be boosted by executing a rapid scan over some portion of the error box.  The optimal values found in our case for a fixed-model type of extrapolation,  0.057 and 0.161 for the baseline and optimistic effective areas, respectively, lead to typical detection timescales of 23 and 8 months.  The values are similar to those found for the fixed model with {\it Swift}-BAT alerts.  It is not possible to do a bandex-type analysis for these GRBs, as spectral information is not available at the time of writing.  However, we can speculate that such a calculation would likely lead to detection efficiencies that are a factor of 1.5 to 2 lower than for the fixed model, which was the general finding for the BATSE GRB population.  Therefore, we conclude that {\it Fermi}-GBM and {\it Swift}-BAT could give rise to detection rates that are generally the same.  However, we point out that in the case of GBM alerts, the scanning mode necessarily means intermittent exposure on the source with therefore only a partial coverage of the temporal emission of the detected GRBs.

It is also possible that future improvements to real-time trigger analysis of GBM bursts could lead to better localization information.  Significant improvements in this area could eliminate the need for scanning or other means to compensate for position uncertainty.  As a simple test, we can take this possibility to an extreme and examine a case in which all positional uncertainty is removed from GBM alerts.  In this case, the detection efficiency values are 0.19 and 0.45, which give detection rates of 1.8 and 4.2 GRBs yr$^{-1}$, $\sim 3$ times greater than the optimal rates from our scan mode simulation.

%=======================
% 5
\section{Results for specific GRBs}
\label{sec:specgrbs}
%=======================

It is useful to consider the spectrum that might be provided by an actual GRB detection.  In this section we show sample spectra from a few different possible GRBs, which are modeled using the parameters summarized in Table \ref{tab:080916Cmod}.

\subsection{080916C}
\label{sec:080916C}
GRB 080916C was seen on September 16, 2008, by {\it Fermi}-LAT and GBM, soon after the beginning of science operations with the instrument.  This GRB is notable both for its high redshift ($z=4.35$; \citealp{greiner09}) and its extremely high isotropic-equivalent luminosity, $8.8 \times 10^{54}$ ergs, or 4.9 \msolar $c^2$ \citep{abdo09a}.  The finding of $>10 $ GeV emission from this burst can be used to set upper limits on the amount of UV light emitted from star-forming galaxies at high redshift \citep{fermiEBL,gilmorePDR}.  We can therefore consider this GRB as an archetypical example of a bright, high-redshift GRB with a hard spectrum known to extend into the multi-GeV energy range. 

We model the high-energy emission from GRB   \newline 080916C using the parameters shown in Table \ref{tab:080916Cmod}.  As in previous sections, an unbroken intrinsic power law extending to 1 TeV is assumed.  The time-integrated flux and high energy spectrum are found using a time-weighted average of the spectra over GBM and LAT energy ranges as presented in Table 1 of \citet{abdo09a}.
\begin{table}
\centering
\begin{tabular}{@{}lccc}
\hline
Parameter & Value (080916C) & Value (BHLZ) & Value (VA) \\
  \hline
  \hline
 Flux &  $4.88 \times 10^{-3}$ &  $1.0 \times 10^{-2}$ & $8.8 \times 10^{-5}$ \\
  \hline
 $ \Gamma$ & -2.16 & -2.1 & -2.0 \\
  \hline
  T90  & 66 s & 100 s & 50 s \\
  \hline
  Redshift  & 4.35  & 0.5 & 2.14 \\
\hline 
\end{tabular}
\caption{Parameters assumed in modeling the three GRBs described in \S \ref{sec:specgrbs}.  These include the time-integrated flux normalization at 1 GeV (with units of GeV$^{-1}$ cm$^{-2}$), the spectral index in $dN/dE$, the T90 duration, and redshift.  Columns show parameters for GRB 080916C (\S \ref{sec:080916C}), a ``bright, hard, low-$z$'' GRB (\S \ref{sec:bhlz}), and a ``very average'' GRB with parameters selected from the medians of the fixed sample used in this paper (\S \ref{sec:va}).
}
\label{tab:080916Cmod}
\end{table}

\begin{figure}
\psfig{file=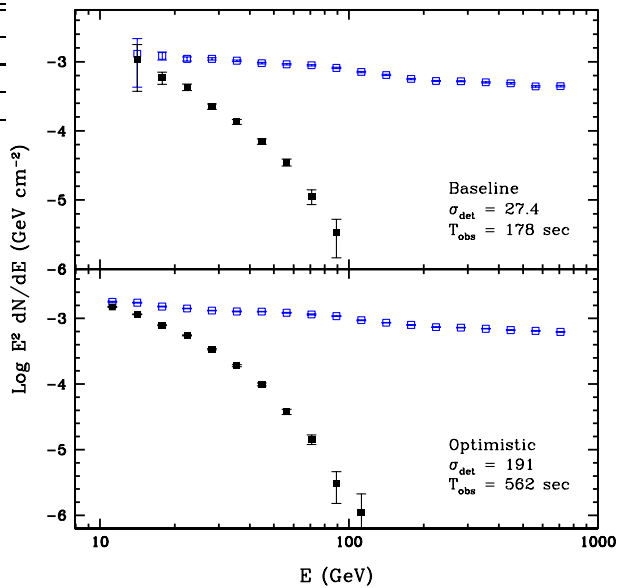,width=\figwidth}
\caption{A simulated realization of the detected spectrum from GRB080916C, assuming the parameters of Table \ref{tab:080916Cmod} and an observation at 20 degrees from zenith.  The top panel is for the baseline effective area function, and the bottom is for the optimistic function.  The blue/grey points refer to the observed spectrum without any attenuation from the EBL.  Black points are after applying the gamma-ray opacity of \citet{gspd11}.  Error bars shown only consider Poisson error in each bin.}
\label{fig:080916Cspect}
\end{figure}

As in \S \ref{sec:results}, the GeV lightcurve is assumed to decay as $t^{-1.5}$ after the T90 period, with no spectral evolution.  Following our analysis with these assumptions, we find that GRB 080916C could be detected at an angle from zenith as high as 39 degrees with the baseline effective area, or 58 degrees with the optimistic area function.  In Figure \ref{fig:080916Cspect}, we show the spectrum that could be expected from an observation of the burst at a zenith angle of 20 degrees.
\noindent In this figure, the spectrum is shown with a bin size of 0.1 dex, along with Poisson error bars for the total number of received photons (signal and noise) in each bin.  For the baseline effective area, we find a total of 682 signal photons  received over the optimal integration timescale of 178 seconds.  A signal is seen up to an energy of 90 GeV in each case, beyond which the signal to noise per bin is well below 1.  For the optimistic effective area, the signal extends conclusively down to 10 GeV, and a total of 17950 photons are detected over an optimal timescale of 562 sec.  Note that the normalizations of the points in the two cases do not appear quite the same, because integrated flux over $T_{obs}$ is being shown, and the timescales of integration are different.  The effect of the EBL is easily seen in a comparison between the attenuated and unattenuated spectra, and the GRB signal is discernible even at energies where an attenuation factor $e^{-\tau} \sim 0.01$ affects observations.

\subsection{Bright, hard, low redshift (BHLZ)}

\label{sec:bhlz}
\begin{figure*}
\plottwo{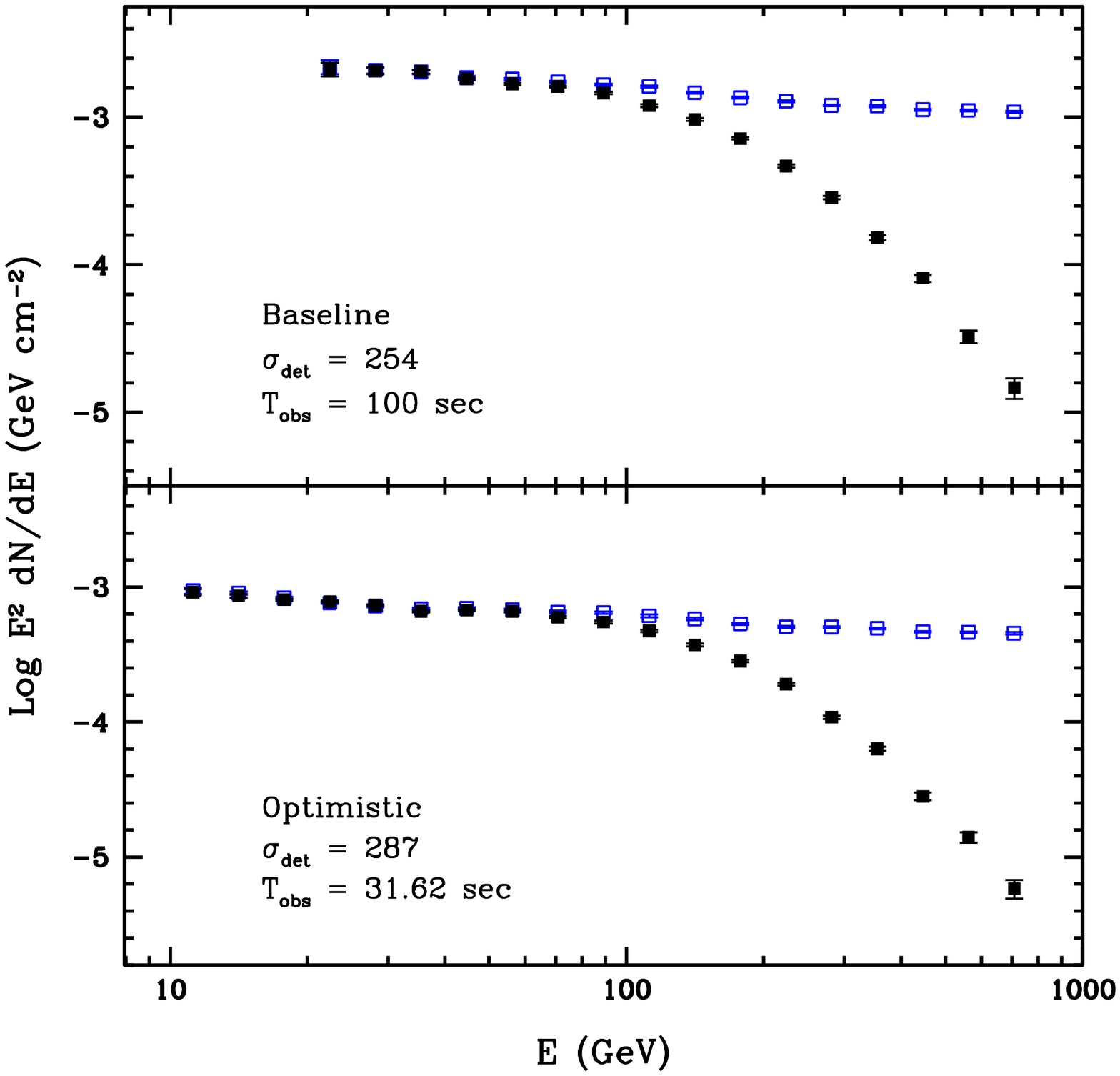}{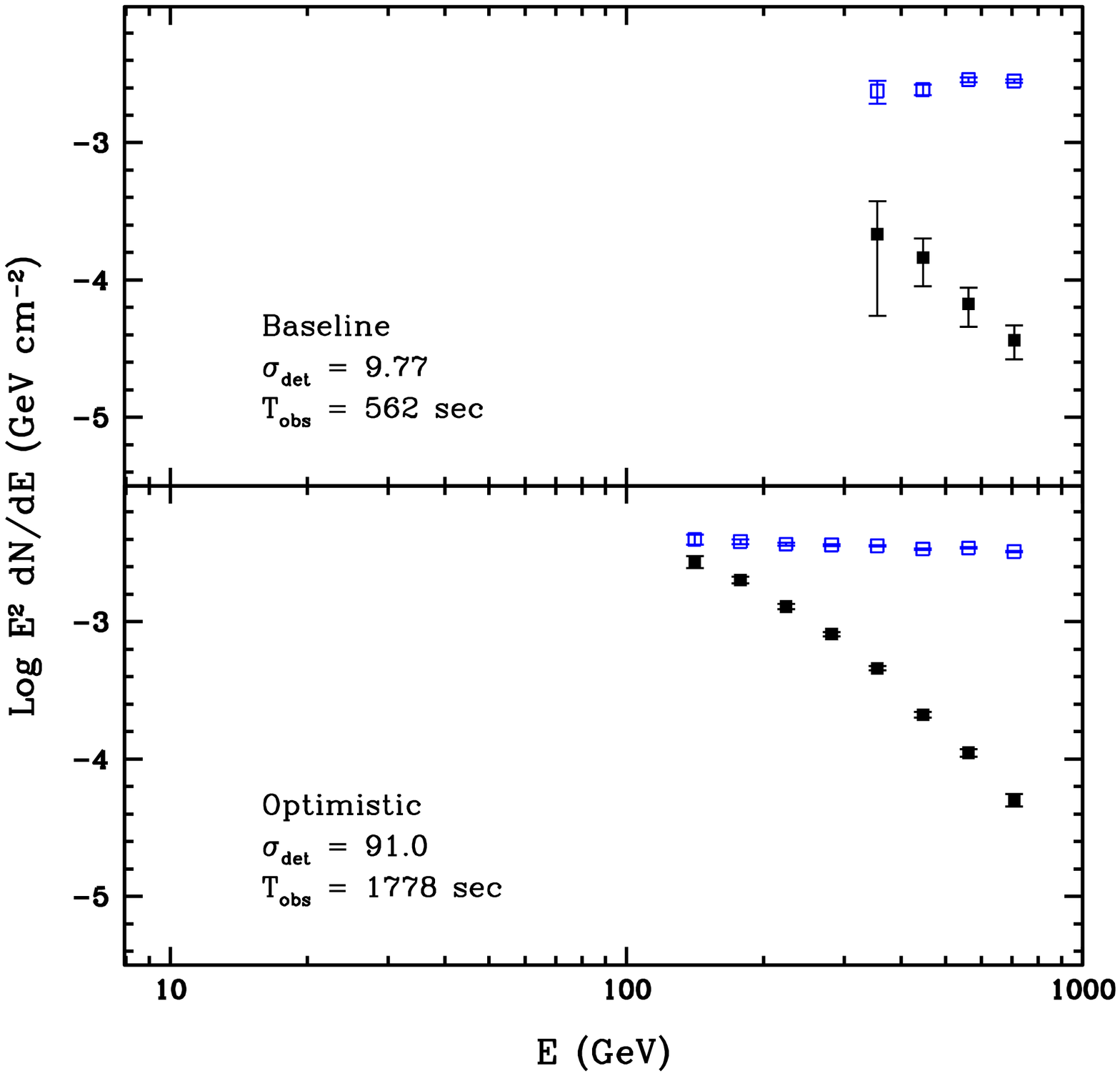}
\caption{A simulated observation of the BHLZ GRB, as described in the text and Table \ref{tab:080916Cmod}.  Spectra are for the burst observed at 30$^\circ$ from zenith (left) and 70$^\circ$ (right).  Point types are as in Fig. \ref{fig:080916Cspect}.  Error bars shown only consider Poisson error in each bin.  Note that the normalizations between the effective area functions are different because the optimal timescale of integration ($T_{obs}$) is found to be different in each case.}
\label{fig:bhlzspect}
\end{figure*}

We now consider the observed spectrum for a bright GRB observed at low redshift.  For this ``BHLZ'' burst, we assume the parameters in the appropriate column of Table \ref{tab:080916Cmod}.  The very high GeV normalization of this GRB (about the 99th percentile in our bandex sample) and its low redshift mean that it can be conclusively detected even at very large angles from zenith.  In Fig. \ref{fig:bhlzspect}, we show an observation of this GRB at an intermediate zenith angle (30$^\circ$) and a very large angle (70$^\circ$).  In the second case, the energy threshold of the telescope is increased by a factor $\cos(70)^{-3} \approx 25$, and therefore the observation is limited to energies above 100 GeV.

\subsection{A ``very average'' (VA) GRB}
\label{sec:va}
\begin{figure}
\psfig{file=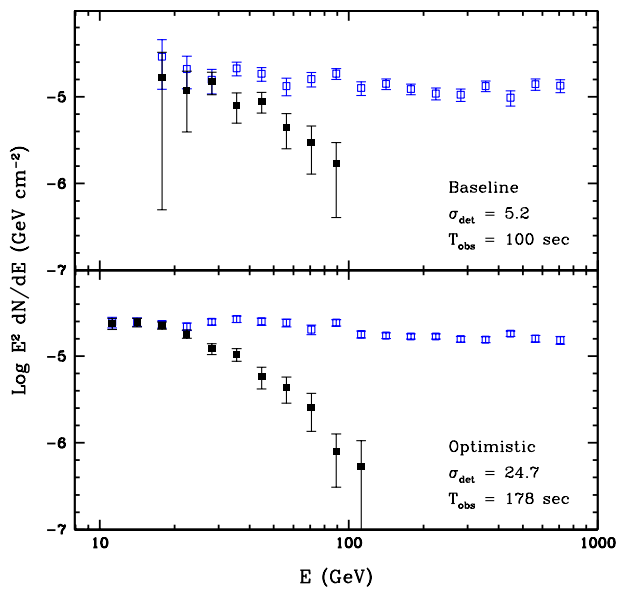,width=\figwidth}
\caption{Simulated spectra from an average GRB, as described in Table \ref{tab:080916Cmod}.  Point types are as in Fig. \ref{fig:080916Cspect}.
}
\label{fig:ea10spect}
\end{figure}

Finally, we repeat our analysis for a GRB that has a redshift, T90 duration, and high energy fluence chosen from the mean values of the fixed sample of GRBs (see Fig. \ref{fig:ctaflu} in Appendix \ref{sec:otherprops}).  Properties are summarized the last column of Table \ref{tab:080916Cmod}.  Such a GRB is found to be only marginally detectable even occurring near zenith with a baseline telescope effective area.  In Fig. \ref{fig:ea10spect}, we show the spectra from such a GRB.

%=======================
% 6
\section{Conclusions}
\label{sec:disc}
%=======================

In this work, we have attempted to make realistic predictions of the GRB detectability with the CTA experiment, in particular by considering a reasonable range of possibilities for the GRB emission and the CTA response functions.  The basic conclusions of this work can be summarized as follows:

\begin{itemize}

\item CTA can be expected to conclusively detect one GRB every 20 to 30 months assuming a baseline effective area and background rate, or 1 to 2 GRBs per year with the optimistic instrument model.

\vspace{0.3cm}

\item Detected GRBs will be at a median redshift $\sim$1, and a typical GRB detected with CTA will provide hundreds or thousands of signal events, mostly appearing below 100 GeV.

\vspace{0.3cm}

\item Intrinsic spectral cutoffs will have little effect on our results provided that they are above an observed energy of 100 GeV ($\sim 200$ GeV intrinsic).

\vspace{0.3cm}

\item  The detection rate of GRBs is a strong function of the instrument energy threshold, and a somewhat weaker function of the typical response time.

\vspace{0.3cm}

\item Follow-up of GRB alerts from {\it Fermi}-GBM could benefit greatly from a scanning-type observation over the GBM error box, which can boost the detection rate by 1.5 to 2 compared to a static observation.  With such a strategy, the detection rate of {\it Fermi}-GBM bursts can be comparable to that of {\it Swift}-detected GRBs.

\vspace{0.3cm}
\item Bright GRBs will provide well-determined spectral information over at least a decade in energy, and will be a valuable source of information about the VHE emission mechanism and intervening cosmological radiation fields.

\end{itemize}

The detection rates we have determined are roughly in agreement with the independent estimate by \citet{kakuwa11} that has been performed concurrently with our own work.
Our findings are contingent on certain assumptions: namely that a satellite instrument (i.e., {\it Swift, SVOM,} or {\it Fermi}-GBM) will be available to provide burst alerts during CTA operations at a rate similar to that seen in recent experience, and that these alerts will be promptly transmitted and followed up  with an instrument slew, when possible.  These rates would approximately double in the case that two such satellites are available.

Our predictions  rely on a number of very uncertain assumptions about gamma-ray bursts that must be determined from limited data.  These include the extrapolation of the spectrum to high energy, the typical lightcurve of the high-energy component, and the amount of extragalactic background light which impedes observations of extragalactic sources in the GeV band.  In the case of the first two, we have been guided by the observation by {\it Fermi}-LAT of $\gtrsim 10$ GeV photons from 4 bright GRBs (080916C, 090510, 090902B, and 090926A).  A constant danger in this model is that these GRBs may not be representative of the population as a whole.  If, in actuality, only a small fraction of GRBs have spectra that continue into the multi-GeV range, then our results here would overestimate the detection rate with CTA by a large factor.  The aforementioned bright LAT-detected GRBs comprise less than 1 percent of the total observed by GBM and it is quite possible that spectral cutoffs routinely exist between the GBM/BATSE energy ranges and the 20--100 GeV band where most GRB photons would be detected (Figure \ref{fig:photcnt}).  Our bandex model incorporates some aspects of a cutoff for a significant fraction of GRBs; those bursts with $\beta$ parameter $\lesssim -3.0$ have much less power in the GeV band than near the Band function peak, and are generally not detectable (see Fig.~\ref{fig:ctabeta} in Appendix \ref{sec:otherprops}).  Over one-third (36 percent) of GRBs in our model fall into this soft category.  But given the limited energy range of BATSE, this number is may not represent the full number of GRBs with spectral turnovers or cutoffs that are below the CTA energy coverage.

Spectral turnovers or cutoffs could exist in GRB spectra due to internal absorption of gamma rays by source photons, or Klein-Nishina suppression of high-energy inverse-Compton emission that could be the basis for GeV-scale emission.  As discussed in \citet{baring06}, internal absorption will lead to a spectral cutoff above an energy determined by the source bulk Lorentz factor ($\Gamma$).  For $\Gamma \gtrsim 1000$ our results are likely unaffected.  In general, only lower limits on $\Gamma$ are available for GRBs; one possible exception being the bright LAT-detected GRB 090926A, where the claimed turnover in the GRB spectrum \citep{bregeon11} can be interpreted as the effect of internal pair opacity \citep{ackermann11}, and used to set limits on the bulk Lorentz factor: $200 \lesssim \Gamma \lesssim 700$.  Lower limits inferred from other bright LAT-detected GRBs are $\Gamma \gtrsim 900$, 1200, and 1000 for GRBs 080916C, 090510, and 090202B, respectively  \citep{ackermann11}.  The distribution in $\Gamma$ is generally unknown for dimmer GRBs, and it is possible that many bursts have factors in the hundreds, rather than thousands.  However, if it is the case that the brightest $\sim 20$ percent of GRBs tend to have unattenuated emission up to $>100$ GeV, then our results would not be strongly affected, as these are the events most likely to be detected in our model.

It is unlikely that our results underestimate the CTA detection rate.  An underestimate of brightness in the GeV band would entail an increase in the typical MeV--GeV brightness ratio (Fig. \ref{fig:flucomp}) above that seen for bright, hard GRBs with a LAT detection.  The lack of LAT detections for most GRBs would seem to disfavor the possibility of large GeV-MeV ratios in fainter GRBs.  Another possibility is that the EBL attenuation of the gamma-ray signal due to UV light for high-redshift GRBs is overestimated in the fiducial model of \citet{gspd11}.  We find that reducing the EBL flux to that of a minimal model, such as G09 Low in Table \ref{tab:eblimpact}, could increase detection rates by as much as 30 to 40 percent, however disagreements between the G09 Low model and the bulk of the high-redshift data make this an unlikely possibility.  Finally, we could have underestimated the performance of CTA itself, though the inclusion of our optimistic effective area model, with its sensitivity at energies as low at $\sim 10$ GeV and reduced background rate, is intended to be as hopeful as reasonably possible about the instrument capabilities.   It is worth noting that CTA will produce a number of marginal detections, with $2 < \sigma <5$.  These are produced at roughly half the rate of bona fide ($>5\sigma$) detections in our models (Fig. \ref{fig:ctasigma}).  These marginal detections would have higher significance than any of the GRB observations in the VERITAS analysis (Section \ref{sec:verupper}; \citealp{aune11}), where the highest significance quoted is 1.8. 

A basic assumption that we have made in this work is that GRBs to which the telescope is triggered, presumably by a satellite instrument, will provide detections at a much higher rate than serendipitous detections, in which a GRB occurs in the field of view of the instrument during other observations.  Simple geometry shows that this is always the case.  The sky coverage fraction of the LSTs, assuming a 4.25 degree field of view and, neglecting edge effects, is about 0.14 percent.  If we extrapolate from the GBM burst rate, and assume $\sim 500$ GRBs occurring per year over all sky, and include the duty cycle factor of 0.1, then we find 1 GRB inside the LST field of view during operations every 13 years.  For the MSTs, with an 8 degree field of view, we have one GRB every 4 years.  Even with the enhanced detection efficiency that is possible with a delay time of zero (Fig. \ref{fig:parvar}; upper left panel), the detection rate for GRBs is still much less for the serendipitous case than for triggered operations.  However, a serendipitous detection over the lifetime of CTA is possible, and would certainly be of great value in constraining the prompt VHE emission of the GRB.

In \S \ref{sec:gbm}, we discuss the prospects for detecting GRB afterglows from {\it Fermi}-GBM alerts.  GBM bursts are found to be more than twice as bright on average than those from {\em Swift}, and a cursory look at the small number of redshifts available for this population suggests that they are typically slightly closer as well. The primary difficulty of detecting GBM bursts lies in the large positional uncertainty of the instrument; only about 10 percent of GRBs will fall in view of the LST for an observation at the center of the GBM error box, for a 4.25 degree field of view.  We show in \S \ref{sec:scanmode} how a scanning mode observation over the error box can improve the detection rate by $\gtrsim 50$ percent.  With this change, we find that the detection rates of GBM bursts with CTA are nearly the same as {\em Swift} GRBs.  If improvements to the GBM angular resolution are possible before the onset of CTA, then our results could be enhanced by a significant factor, as described in Fig.~\ref{fig:gbmerrred}.  We have considered the possibility of improved localization with the use of a simple linear scaling of the positional error in this figure, as it is difficult to quantify to what extent such an error reduction could take place over the next several years.  It is worth emphasizing that because the brightest GRBs are generally those with the highest probability of detection, reductions in the errors for especially bright and/or hard GRBs will be the most advantageous in increasing the detection efficiency for GBM alerts.  

Our results show that GRB detection with CTA will rely heavily on the sensitivity achievable in the 20 to 100 GeV band.  As shown in Fig.~\ref{fig:photcnt}, only about 20 percent of the gamma rays found for typical detected GRB are at energies $> 100$ GeV; this is despite the large upturn of the effective area function in both of our assumed telescope models (Fig.~\ref{fig:ea}) at this energy.  It should be emphasized that detection of GRBs with CTA is therefore heavily reliant on the performance of the LSTs.   It should also be noted that without an LST component, we find that the detection rate for CTA is only marginally higher than for the VERITAS instrument (Table \ref{tab:deinstsum}), and is less than one-third the rate for the complete CTA instrument.  The importance of the LSTs is even greater if there exist intrinsic spectral cutoffs in GRB spectra at energies between 20 and 100 GeV.  We note, however, that MSTs could play a significant role in providing high statistics data above 100 GeV to perform time-resolved spectroscopy which could reveal much about the GRB physics.

An exciting prospect is to use the spectrum of a GRB seen by CTA to constrain the EBL.  While gamma-ray sources have been used in many cases to help constrain the EBL (see references in \citealp{gspd11}), these attempts have mainly focused on relatively low redshift blazars.  A high-statistics GRB detection by CTA at $z>1$ could greatly improve our understanding of how the EBL evolves with redshift.  Because the EBL impact is significant below 100 GeV at these distances, the LSTs will be crucial for such science.  One great advantage of CTA is its ability to potentially detect simultaneously both the attenuated and unabsorbed portions of a gamma-ray spectrum, which allows much more robust limits on the EBL than if only the attenuated spectrum is seen and the intrinsic spectrum must be derived theoretically \citep{raue&mazin10}.  For GRBs at and above the median redshift of $z=1.7$, an energy threshold of $\lesssim 20$ GeV will be needed to effectively capture the unattenuated slope.

%===================================
\section*{Acknowledgments}

This work has been supported by a SISSA postdoctoral fellowship (RCG) and grants from the Fermi Guest Investigator Program and the US National Science Foundation.   The authors thank the VERITAS Collaboration for the use of unpublished results from the detector Monte Carlo simulation.  They also thank Taylor Aune for providing early access to the VERITAS limits on GRB fluence, and acknowledge the GBM operations team for continued access to prompt burst locations and for the GRB catalog information used in the predictive calculations presented here.  RCG  also thanks Vladimir Vassiliev for a useful discussion related to CTA performance, and Susumu Inoue, Jun Kakuwa, and Ryo Yamazaki for helpful discussions concerning this calculation.

%=====================================

%\bibliographystyle{spbasic} 
%\bibliography{../../writings/eblpaper}

\begin{thebibliography}{60}
\providecommand{\natexlab}[1]{#1}
\providecommand{\url}[1]{{#1}}
\providecommand{\urlprefix}{URL }
\expandafter\ifx\csname urlstyle\endcsname\relax
  \providecommand{\doi}[1]{DOI~\discretionary{}{}{}#1}\else
  \providecommand{\doi}{DOI~\discretionary{}{}{}\begingroup
  \urlstyle{rm}\Url}\fi
\providecommand{\eprint}[2][]{\url{#2}}

\bibitem[{{Abdo} et~al(2007){Abdo}, {Allen}, {Berley}, {Blaufuss}, {Casanova},
  {Dingus}, {Ellsworth}, {Gonzalez}, and {Goodman}}]{abdo07}
{Abdo} AA, et al.: {Milagro Constraints on
  Very High Energy Emission from Short-Duration Gamma-Ray Bursts}. \apj~ 666, 361--367 (2007)

\bibitem[{{Abdo} et~al(2009{\natexlab{a}}){Abdo}, {Ackermann}, {Ajello},
  {Asano}, {Atwood}, {Axelsson}, {Baldini}, {Ballet}, and
  {Barbiellini}}]{abdo09c}
{Abdo} AA, et al.: {Fermi
  Observations of GRB 090902B: A Distinct Spectral Component in the Prompt and
  Delayed Emission}. \apjl~706, L138--L144 
  (2009a)  

\bibitem[{{Abdo} et~al(2009{\natexlab{b}}){Abdo}, {Ackermann}, {Ajello},
  {Asano}, {Atwood}, {Axelsson}, {Baldini}, {Ballet}, {Barbiellini}, {Baring},
  and et~al.}]{abdo09d}
{Abdo} AA, et al.: {A limit on the variation of the speed of light arising
  from quantum gravity effects}. \nat~462, 331--334 (2009{\natexlab{b}})

\bibitem[{{Abdo} et~al(2009{\natexlab{c}}){Abdo}, {Ackermann}, {Ajello},
  {Asano}, {Atwood}, {Axelsson}, {Baldini}, {Ballet}, {Barbiellini}, {Baring},
  and {Bastieri}}]{abdo10a}
{Abdo} AA, et al.:
  {Fermi Observations of GRB 090902B: A Distinct Spectral
  Component in the Prompt and Delayed Emission}. \apjl~706, L138--L144 (2009{\natexlab{c}})

\bibitem[{{Abdo} et~al(2009{\natexlab{d}})}]{abdo09a}
  {Abdo}, et al.: {Fermi Observations
  of High-Energy Gamma-Ray Emission from GRB 080916C}. Science 323, 1688 (2009{\natexlab{d}})

\bibitem[{{Abdo} et~al(2010){Abdo}, {Ackermann}, {Ajello}, {Allafort},
  {Atwood}, {Baldini}, {Ballet}, {Barbiellini}, {Baring}, {Bastieri},
  {Baughman}, {Bechtol}, {Bellazzini}, {Berenji}, and {Bhat}}]{fermiEBL}
{Abdo} AA,et al.: {Fermi Large Area
  Telescope Constraints on the Gamma-ray Opacity of the Universe}. \apj~723, 1082--1096 (2010)

\bibitem[{{Acciari} et~al(2011){Acciari}, {Aliu}, {Arlen}, {Aune}, {Beilicke},
  {Benbow}, {Bradbury}, {Buckley}, {Bugaev}, {Byrum}, {Cannon}, {Cesarini},
  {Christiansen}, {Ciupik}, and {Collins-Hughes}}]{aune11}
{Acciari} VA, et al.: {VERITAS
  Observations of Gamma-Ray Bursts Detected by Swift}. \apj~743, 62 (2011)

\bibitem[{{Ackermann}(2010)}]{ackermann10}
{Ackermann} M: {Fermi Observations of GRB 090510: A Short Hard Gamma-Ray
  Burst with an Additional, Hard Power-Law Component from 10 keV to GeV
  Energies}. ArXiv:1005.2141 (2010) 

\bibitem[{{Ackermann} et~al(2011){Ackermann}, {Ajello}, {Asano}, {Axelsson},
  {Baldini}, {Ballet}, {Barbiellini}, {Baring}, {Bastieri}, {Bechtol},
  {Bellazzini}, {Berenji}, {Bhat}, {Bissaldi}, and {Blandford}}]{ackermann11}
{Ackermann} M, et al.: {Detection of a
  Spectral Break in the Extra Hard Component of GRB 090926A}. \apj~729, 114 (2011) 

\bibitem[{{Aharonian} et~al(2009){Aharonian}, {Akhperjanian}, {Barres de
  Almeida}, {Bazer-Bachi}, {Behera}, {Benbow}, {Bernl{\"o}hr}, {Boisson},
  {Bochow}, {Borrel}, {Braun}, {Brion}, {Brucker}, {Brun}, and
  {B{\"u}hler}}]{aharonian09}
{Aharonian} F, et al.: {HESS
  observations of {$\gamma$}-ray bursts in 2003-2007}. \aap~495, 505--512 (2009)
 

\bibitem[{{Albert} et~al(2007){Albert}, {Aliu}, {Anderhub}, {Antoranz},
  {Armada}, {Baixeras}, {Barrio}, {Bartko}, {Bastieri}, {Becker}, {Bednarek},
  {Berger}, {Bigongiari}, {Biland}, {Bock}, and {Bordas}}]{albert07d}
{Albert} J, et al.: {MAGIC Upper Limits
  on the Very High Energy Emission from Gamma-Ray Bursts}. \apj~667, 358--366 (2007)
 

\bibitem[{{Albert} et~al(2008){Albert}, {Aliu}, {Anderhub}, {Antoranz},
  {Armada}, {Baixeras}, {Barrio}, {Bartko}, {Bastieri}, {Becker}, {Bednarek},
  and {Berger}}]{albert08a}
{Albert} J, et al.:  {VHE {$\gamma$}-Ray Observation of the Crab Nebula and its Pulsar with
  the MAGIC Telescope}. \apj~674, 1037--1055 (2008)

\bibitem[{{Asano} et~al(2010){Asano}, {Inoue}, and
  {M{\'e}sz{\'a}ros}}]{asano10}
{Asano} K, {Inoue} S, {M{\'e}sz{\'a}ros} P: {Prompt X-ray and Optical
  Excess Emission Due to Hadronic Cascades in Gamma-ray Bursts}. \apjl~725, L121--L125 (2010) 


\bibitem[{{Atkins} et~al(2003){Atkins}, {Benbow}, {Berley}, {Chen}, {Coyne},
  {Dingus}, {Dorfan}, {Ellsworth}, {Evans}, {Falcone}, and
  {Fleysher}}]{atkins03}
{Atkins} R, et al.: {The
  High-Energy Gamma-Ray Fluence and Energy Spectrum of GRB 970417a from
  Observations with Milagrito}. \apj~583, 824--832 (2003)

\bibitem[{{Atwood} et~al(2009){Atwood}, {Abdo}, {Ackermann}, {Althouse},
  {Anderson}, {Axelsson}, {Baldini}, {Ballet}, {Band}, {Barbiellini},
  {Bartelt}, and {Bastieri}}]{atwood09}
{Atwood} WB, et al.: {The Large Area Telescope on the Fermi Gamma-Ray Space
  Telescope Mission}. \apj~697, 1071--1102 (2009)

\bibitem[{{Band} et~al(1993){Band}, {Matteson}, {Ford}, {Schaefer}, {Palmer},
  {Teegarden}, {Cline}, {Briggs}, {Paciesas}, {Pendleton}, and
  {Fishman}}]{band93}
{Band} D, et al.: {BATSE
  observations of gamma-ray burst spectra. I - Spectral diversity}. \apj~413, 281--292 (1993) 

\bibitem[{{Band} et~al(2009){Band}, {Axelsson}, {Baldini}, {Barbiellini},
  {Baring}, {Bastieri}, {Battelino}, {Bellazzini}, {Bissaldi}, {Bogaert},
  {Bonnell}, {Chiang}, {Cohen-Tanugi}, {Connaughton}, {Cutini}, {de Palma},
  {Dingus}, and {do Couto e Silva}}]{band09}
{Band} DL, et al.: {Prospects for GRB Science with the
  Fermi Large Area Telescope}. \apj~701, 1673--1694 (2009)

\bibitem[{{Baring}(2006)}]{baring06}
{Baring} MG: {Temporal Evolution of Pair Attenuation Signatures in
  Gamma-Ray Burst Spectra}. \apj~650, 1004--1019  (2006)

\bibitem[{{Bastieri} et~al(2005){Bastieri}, {Galante}, {Gaug}, {Garczarczyk},
  {Longo}, {Mizobuchi}, and {Peruzzo}}]{bastieri05}
{Bastieri} D et al.: {The MAGIC Telescope and the observation of Gamma Ray
  Bursts}. Geophysics Space Physics C 28, 711 (2005) 

\bibitem[{{Bissaldi}(2011)}]{bissaldi11}
{Bissaldi} E: {The Fermi Gamma-ray Burst Monitor: Results from the first
  two years}. ArXiv:1101.3697 (2011)

\bibitem[{{Bregeon} et~al(2011){Bregeon}, {Goldstein}, {Preece}, {Takahashi},
  {Toma}, and {Uehara}}]{bregeon11}
{Bregeon} J, {Goldstein} A, {Preece} R, {Takahashi} H, {Toma} K, {Uehara} T:
 {Detection of a spectral break in the extra hard component of GRB
  090926A}. ArXiv:1101.2082   (2011)

\bibitem[{{Dermer}(2010)}]{dermer10}
{Dermer} CD: {First Light on GRBs with Fermi}. In {N~Kawai \&
  S~Nagataki} (ed) American Institute of Physics Conference Series, American
  Institute of Physics Conference Series, vol 1279, pp 191--199 (2010)

\bibitem[{{Dingus}(1995)}]{dingus95}
{Dingus} BL: {EGRET Observations of {$\geq$} 30 MeV Emission from the
  Brightest Bursts Detected by BATSE}. \apss~231, 187--190 (1995)

\bibitem[{{Fan} et~al(2008){Fan}, {Piran}, {Narayan}, and {Wei}}]{fan08}
{Fan} YZ, {Piran} T, {Narayan} R, {Wei} DM: {High-energy afterglow
  emission from gamma-ray bursts}. \mnras~384, 1483--1501  (2008)

\bibitem[{{Franceschini} et~al(2008){Franceschini}, {Rodighiero}, and
  {Vaccari}}]{franceschini08}
{Franceschini} A, {Rodighiero} G, {Vaccari} M: {Extragalactic
  optical-infrared background radiation, its time evolution and the cosmic
  photon-photon opacity}. \aap~487, 837--852  (2008)

\bibitem[{{Gao} et~al(2009){Gao}, {Mao}, {Xu}, and {Fan}}]{gao09}
{Gao} W, {Mao} J, {Xu} D, {Fan} Y: {GRB 080916C and GRB 090510: The
  High-Energy Emission and the Afterglow}. \apjl~706, L33--L36 (2009)

\bibitem[{{Garczarczyk} et~al(2008){Garczarczyk}, {Antonelli}, {La Barbera},
  {Bastieri}, {Convino}, {Galante}, {Gaug}, {Longo}, and
  {Scapin}}]{garczarczyk08}
{Garczarczyk} M, et al.: {Observation of gamma ray
  bursts at very high energies with the MAGIC telescope}. In {Huang} YF, {Dai}
  ZG, {Zhang} B (eds) American Institute of Physics Conference Series, American
  Institute of Physics Conference Series, vol 1065, pp 342--344 (2008)

\bibitem[{{Ghirlanda} et~al(2010){Ghirlanda}, {Ghisellini}, and
  {Nava}}]{ghirlanda09}
{Ghirlanda} G, {Ghisellini} G, {Nava} L: {The onset of the GeV afterglow
  of GRB 090510}. \aap~510, L7 (2010)

\bibitem[{{Ghisellini} et~al(2010){Ghisellini}, {Ghirlanda}, {Nava}, and
  {Celotti}}]{ghisellini10}
{Ghisellini} G, {Ghirlanda} G, {Nava} L, {Celotti} A: {GeV emission from
  gamma-ray bursts: a radiative fireball?} \mnras~403, 926--937 (2010)

\bibitem[{{Gilmore}(2011)}]{gilmorePDR}
{Gilmore} RC: {Constraining the near-IR background light from
  Population-III stars using high redshift gamma-ray sources}.  \mnras~420, 800--809 (2011)

\bibitem[{{Gilmore} et~al(2009){Gilmore}, {Madau}, {Primack}, {Somerville}, and
  {Haardt}}]{gilmoreUV}
{Gilmore} RC, {Madau} P, {Primack} JR, {Somerville} RS, {Haardt} F: {GeV
  gamma-ray attenuation and the high-redshift UV background}. \mnras~399, 1694--1708  (2009)

\bibitem[{{Gilmore} et~al(2010){Gilmore}, {Prada}, and {Primack}}]{gilmoreGRB}
{Gilmore} RC, {Prada} F, {Primack} J: {Modelling gamma-ray burst
  observations by Fermi and MAGIC including attenuation due to diffuse
  background light}. \mnras~402, 565--574 (2010)

\bibitem[{{Gilmore} et~al(2011){Gilmore}, {Somerville}, {Primack}, and
  {Dom{\'{\i}}nguez}}]{gspd11}
{Gilmore} RC, {Somerville} RS, {Primack} JR, {Dom{\'{\i}}nguez} A: 
  {Semi-analytic modeling of the EBL and consequences for extragalactic
  gamma-ray spectra}. MNRAS accepted, ArXiv:1104.0671 (2011)

\bibitem[{{G{\"o}tz} et~al(2009){G{\"o}tz}, {Paul}, {Basa}, {Wei}, {Zhang},
  {Atteia}, {Barret}, {Cordier}, {Claret}, {Deng}, {Fan}, {Hu}, {Huang},
  {Mandrou}, {Mereghetti}, {Qiu}, and {Wu}}]{gotz09}
{G{\"o}tz} D, et al.:  {SVOM: a new mission for Gamma-Ray
  Burst Studies}. In: {C~Meegan, C~Kouveliotou, \& N~Gehrels} (ed) American
  Institute of Physics Conference Series, American Institute of Physics
  Conference Series, vol 1133, pp 25--30  (2009)

\bibitem[{{Greiner} et~al(2009){Greiner}, {Clemens}, {Kr{\"u}hler}, {von
  Kienlin}, {Rau}, {Sari}, {Fox}, {Kawai}, {Afonso}, {Ajello}, {Berger}, and
  {Cenko}}]{greiner09}
{Greiner} J, et al.: 
  {The redshift and afterglow of the extremely energetic gamma-ray burst GRB
  080916C}. \aap~498, 89--94 (2009)

\bibitem[{{Hattori} et~al(2007){Hattori}, {Aoki}, and {Kawai}}]{hattori07}
{Hattori} T, {Aoki} K, {Kawai} N: {GRB 070521: subaru observations and
  possible host detection.} GRB Coordinates Network 6444, 1  (2007)

\bibitem[{{Kakuwa} et~al(2011){Kakuwa}, {Murase}, {Toma}, {Inoue}, {Yamazaki},
  and {Ioka}}]{kakuwa11}
{Kakuwa} J, {Murase} K, {Toma} K, {Inoue} S, {Yamazaki} R, {Ioka} K: 
  {Prospects for Detecting Gamma-Ray Bursts at Very High Energies with the
  Cherenkov Telescope Array}. ArXiv:1112.5940 (2011)

\bibitem[{{Kumar} and {Barniol Duran}(2010)}]{kumar&barniolduran10}
{Kumar} P, {Barniol Duran} R: {External forward shock origin of
  high-energy emission for three gamma-ray bursts detected by Fermi}. \mnras~409, 226--236  (2010)

\bibitem[{{Le} and {Dermer}(2009)}]{le&dermer09}
{Le} T, {Dermer} CD: {Gamma-ray Burst Predictions for the Fermi Gamma Ray
  Space Telescope}. \apj~700, 1026--1033 (2009)

\bibitem[{{Li} and {Ma}(1983)}]{li&ma83}
{Li} T, {Ma} Y: {Analysis methods for results in gamma-ray astronomy}.
  \apj~272, 317--324 (1983)

\bibitem[{{Lloyd-Ronning} et~al(2002){Lloyd-Ronning}, {Fryer}, and
  {Ramirez-Ruiz}}]{lloydronning02}
{Lloyd-Ronning} NM, {Fryer} CL, {Ramirez-Ruiz} E: {Cosmological Aspects
  of Gamma-Ray Bursts: Luminosity Evolution and an Estimate of the Star
  Formation Rate at High Redshifts}. \apj~574, 554--565 (2002)

\bibitem[{{Madau} and {Phinney}(1996)}]{madau&phinney96}
{Madau} P, {Phinney} ES: {Constraints on the Extragalactic Background
  Light from Gamma-Ray Observations of High-Redshift Quasars}. \apj~456, 124 (1996)

\bibitem[{{Meegan} et~al(2009){Meegan}, {Lichti}, {Bhat}, {Bissaldi}, {Briggs},
  {Connaughton}, {Diehl}, {Fishman}, {Greiner}, {Hoover}, {van der Horst}, {von
  Kienlin}, {Kippen}, {Kouveliotou}, {McBreen}, {Paciesas}, {Preece},
  {Steinle}, {Wallace}, {Wilson}, and {Wilson-Hodge}}]{meegan09}
{Meegan} C, et al.: 
  {The Fermi Gamma-ray Burst Monitor}. \apj~702, 791--804 (2009)

\bibitem[{{Nikishov}(1962)}]{nikishov62}
{Nikishov} AI: {Absorption of High-Energy Photons in the Universe}.
  Soviet Physics JETP 14, 393--394 (1962)

\bibitem[{{Paciesas} et~al(2012){Paciesas}, {Meegan}, {von Kienlin}, {Bhat},
  {Bissaldi}, {Briggs}, {Burgess}, {Chaplin}, {Connaughton}, {Diehl},
  {Fishman}, {Fitzpatrick}, {Foley}, {Gibby}, and {Giles}}]{paciesas12}
{Paciesas} WS, et al.: {The Fermi GBM
  Gamma-Ray Burst Catalog: The First Two Years}. \apjs~199, 18 (2012)

\bibitem[{{Piran}(2004)}]{piran04}
{Piran} T: {The physics of gamma-ray bursts}. Reviews of Modern Physics 76, 1143--1210 (2004)

\bibitem[{{Preece} et~al(2000){Preece}, {Briggs}, {Mallozzi}, {Pendleton},
  {Paciesas}, and {Band}}]{preece00}
{Preece} RD, {Briggs} MS, {Mallozzi} RS, {Pendleton} GN, {Paciesas} WS, {Band}
  DL: {The BATSE Gamma-Ray Burst Spectral Catalog. I. High Time
  Resolution Spectroscopy of Bright Bursts Using High Energy Resolution Data}.
  \apjs~126, 19--36 (2000)

\bibitem[{{Rando}(2009)}]{rando09}
{Rando} R:  {Post-launch performance of the Fermi Large Area Telescope}.
  ArXiv:0907.0626  (2009)

\bibitem[{{Raue} and {Mazin}(2010)}]{raue&mazin10}
{Raue} M, {Mazin} D: {Potential of the next generation VHE instruments to
  probe the EBL (I): The low- and mid-VHE}. Astroparticle Physics 34, 245--256 (2010)

\bibitem[{{Razzaque} et~al(2010){Razzaque}, {Dermer}, and
  {Finke}}]{razzaque09a}
{Razzaque} S, {Dermer} CD, {Finke} JD:  {Synchrotron Radiation from
  Ultra-High Energy Protons and the Fermi Observations of GRB 080916C}. The
  Open Astronomy Journal 3, 150--155  (2010)

\bibitem[{{Romano}(2010)}]{romano10}
{Romano} P: {Swift: the science across the rainbow. Mission Overview and
  Highlights of Results}. ArXiv:1010.2206 (2010)

\bibitem[{{Salvaterra} et~al(2009{\natexlab{a}}){Salvaterra}, {Della Valle},
  {Campana}, {Chincarini}, {Covino}, {D'Avanzo}, {Fern{\'a}ndez-Soto},
  {Guidorzi}, {Mannucci}, {Margutti}, and {Th{\"o}ne}}]{salvaterra09}
{Salvaterra} R, et al.: {GRB090423 at a redshift of
  z{$\sim$}8.1}. \nat~461, 1258--1260 (2009{\natexlab{a}})

\bibitem[{{Salvaterra} et~al(2009{\natexlab{b}}){Salvaterra}, {Guidorzi},
  {Campana}, {Chincarini}, and {Tagliaferri}}]{salvaterra09a}
{Salvaterra} R, {Guidorzi} C, {Campana} S, {Chincarini} G, {Tagliaferri} G:
  {Evidence for luminosity evolution of long gamma-ray
  bursts in Swift data}. \mnras~396, 299--303 (2009{\natexlab{b}})

\bibitem[{{Sari} and {Piran}(1997)}]{sari&piran97}
{Sari} R, {Piran} T: {Cosmological gamma-ray bursts: internal versus
  external shocks}. \mnras~287, 110--116 (1997)

\bibitem[{{Somerville} et~al(2008){Somerville}, {Hopkins}, {Cox}, {Robertson},
  and {Hernquist}}]{somerville08}
{Somerville} RS, {Hopkins} PF, {Cox} TJ, {Robertson} BE, {Hernquist} L:
  {A semi-analytic model for the co-evolution of galaxies, black holes and
  active galactic nuclei}. \mnras~391, 481--506  (2008)

\bibitem[{{Somerville} et~al(2011){Somerville}, {Gilmore}, {Primack}, and
  {Dominguez}}]{sgpd11}
{Somerville} RS, {Gilmore} RC, {Primack} JR, {Dominguez} A: {Galaxy
  Properties from the Ultra-violet to the Far-Infrared: Lambda-CDM models
  confront observations}. MNRAS accepted, ArXiv:1104.0671 (2011)

\bibitem[{{Teshima}(2011)}]{teshima11}
{Teshima} M: {Design Study of a CTA Large Size Telescope (LST)}. In
  Proc. of the 32nd Int. Cosmic Ray Conf., 32ND INTERNATIONAL COSMIC RAY
  CONFERENCE, vol~9, p 149 (2011) 

\bibitem[{{Actis} et~al(2011)}]{ctaconcept10}
{Actis} et al.: {Design concepts for the Cherenkov Telescope Array CTA: an advanced facility for ground-based high-energy gamma-ray astronomy}. Exp. Astro. 32, 193--316  (2011)

\bibitem[{{Wang} et~al(2005){Wang}, {Cheng}, {Dai}, and {Lu}}]{wang05}
{Wang} XY, {Cheng} KS, {Dai} ZG, {Lu} T: {High-energy component of GRB
  941017 revisited and the reverse-shock synchrotron self-Compton emission}.
  \aap~439, 957--961 (2005)

\bibitem[{{Zou} et~al(2009){Zou}, {Fan}, and {Piran}}]{zou09}
{Zou} Y, {Fan} Y, {Piran} T: {The possible high-energy emission from GRB
  080319B and origins of the GeV emission of GRBs 080514B, 080916C and
  081024B}. \mnras~396, 1163--1170 (2009)

\end{thebibliography}

%\pagebreak[4]
\appendix

\section[]{Other properties of detected GRBs}
\label{sec:otherprops}

\begin{figure*}
\plottwo{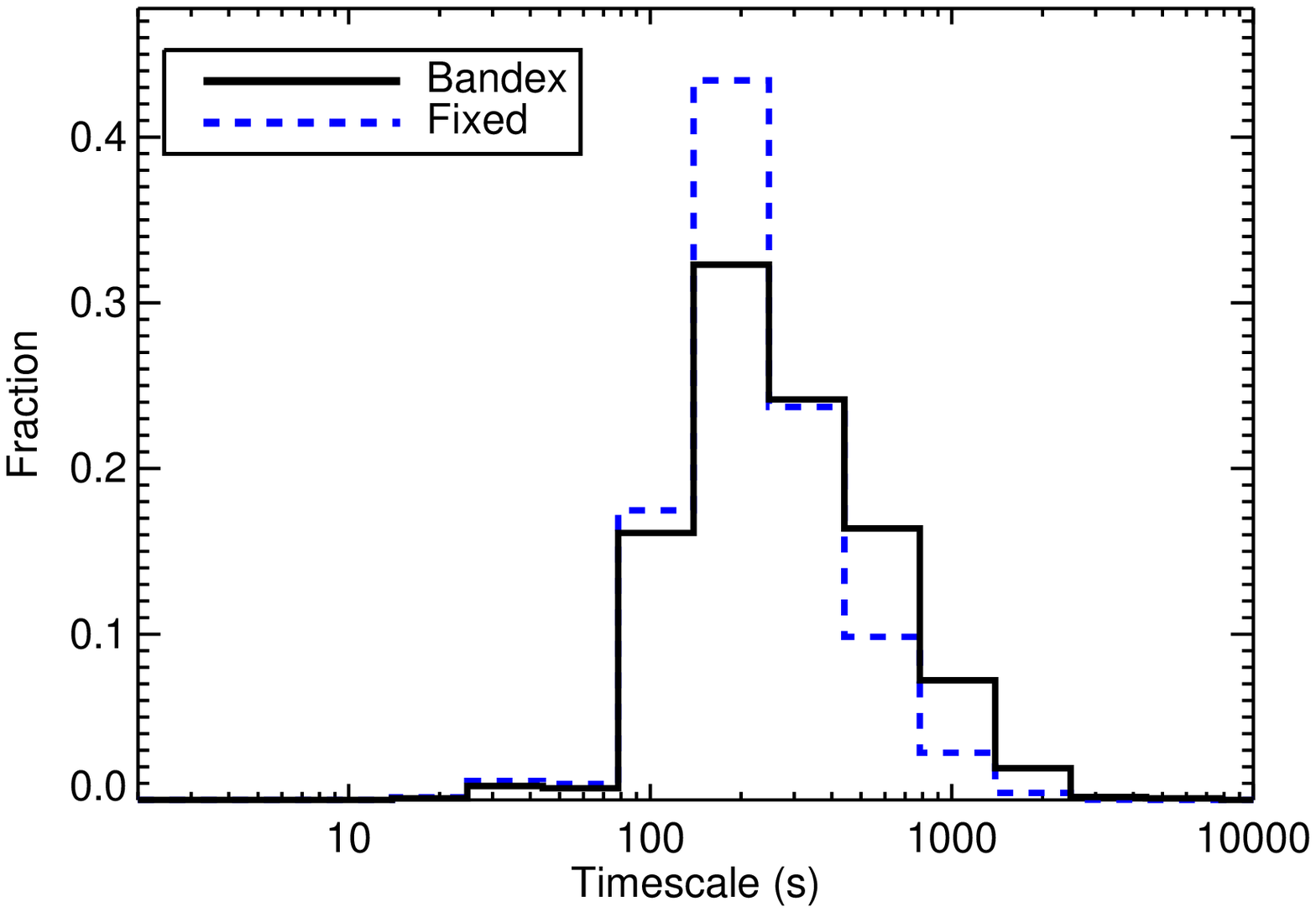}{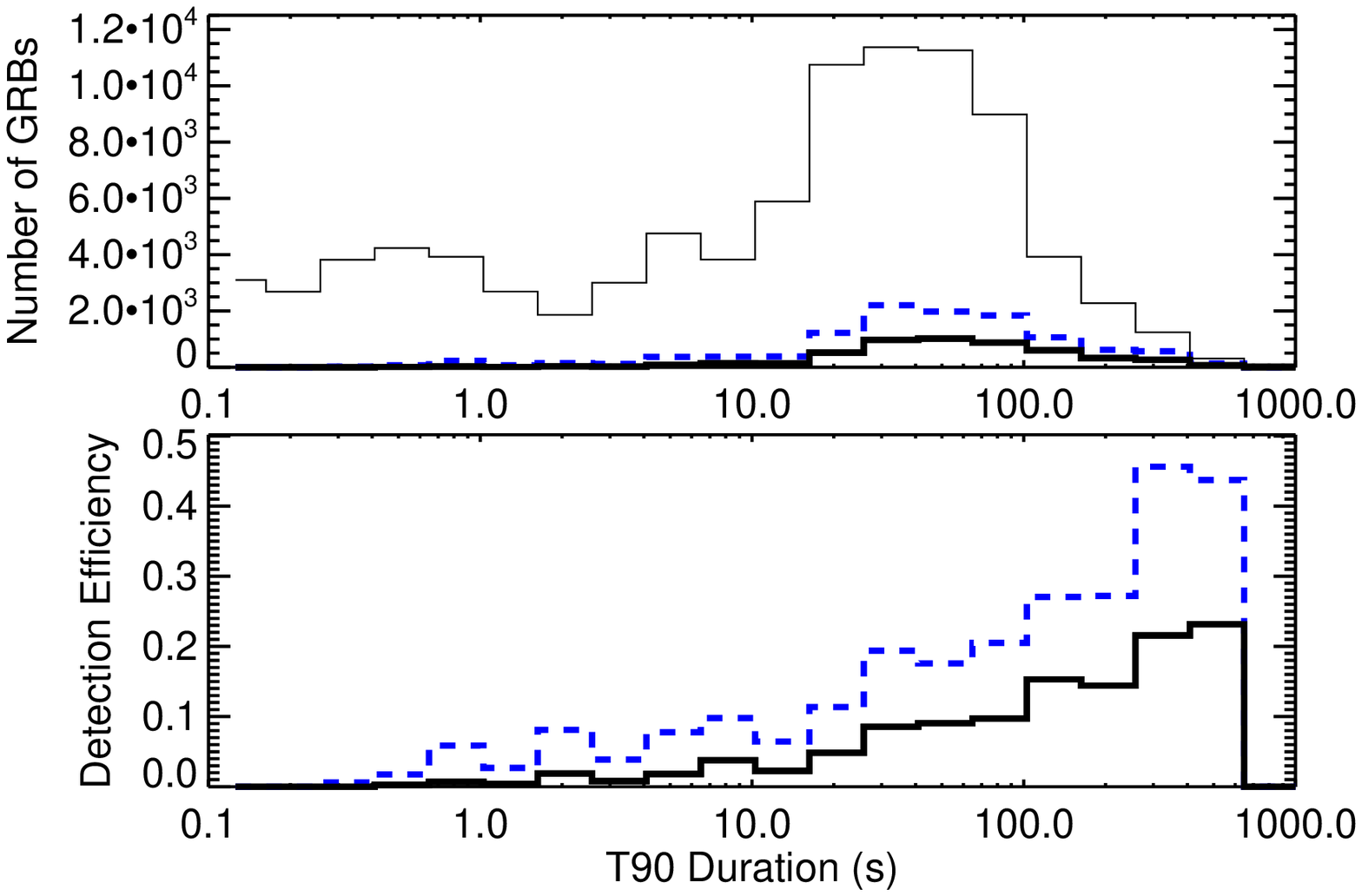}
\caption{{\bf Left:} The distribution of integration timescales that maximize detection significance for detected GRBs, for the bandex (solid black) and fixed (broken blue) models.  {\bf Right:} Comparison of T90 for detected GRBs with the whole population, for $3\times 10^4$ simulated GRBs.  In the top panel, the thin line is the distribution of the full population, and the solid black and broken blue lines are the number of detected GRBs for the bandex and fixed models, respectively.  The bottom panel shows the fraction of GRBs detected in each bin.}
\label{fig:timescales}
\end{figure*}

\begin{figure*}
\plottwo{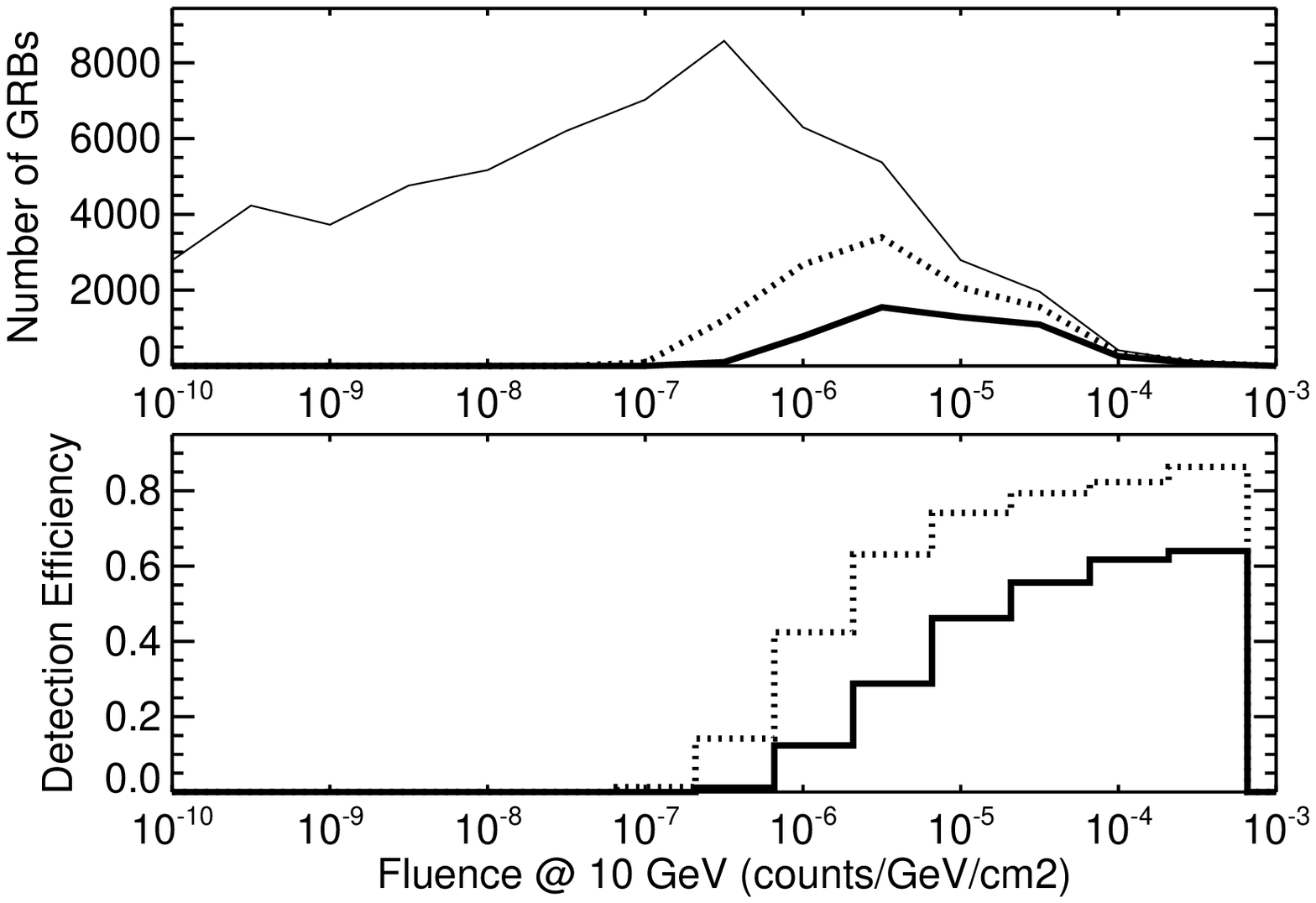}{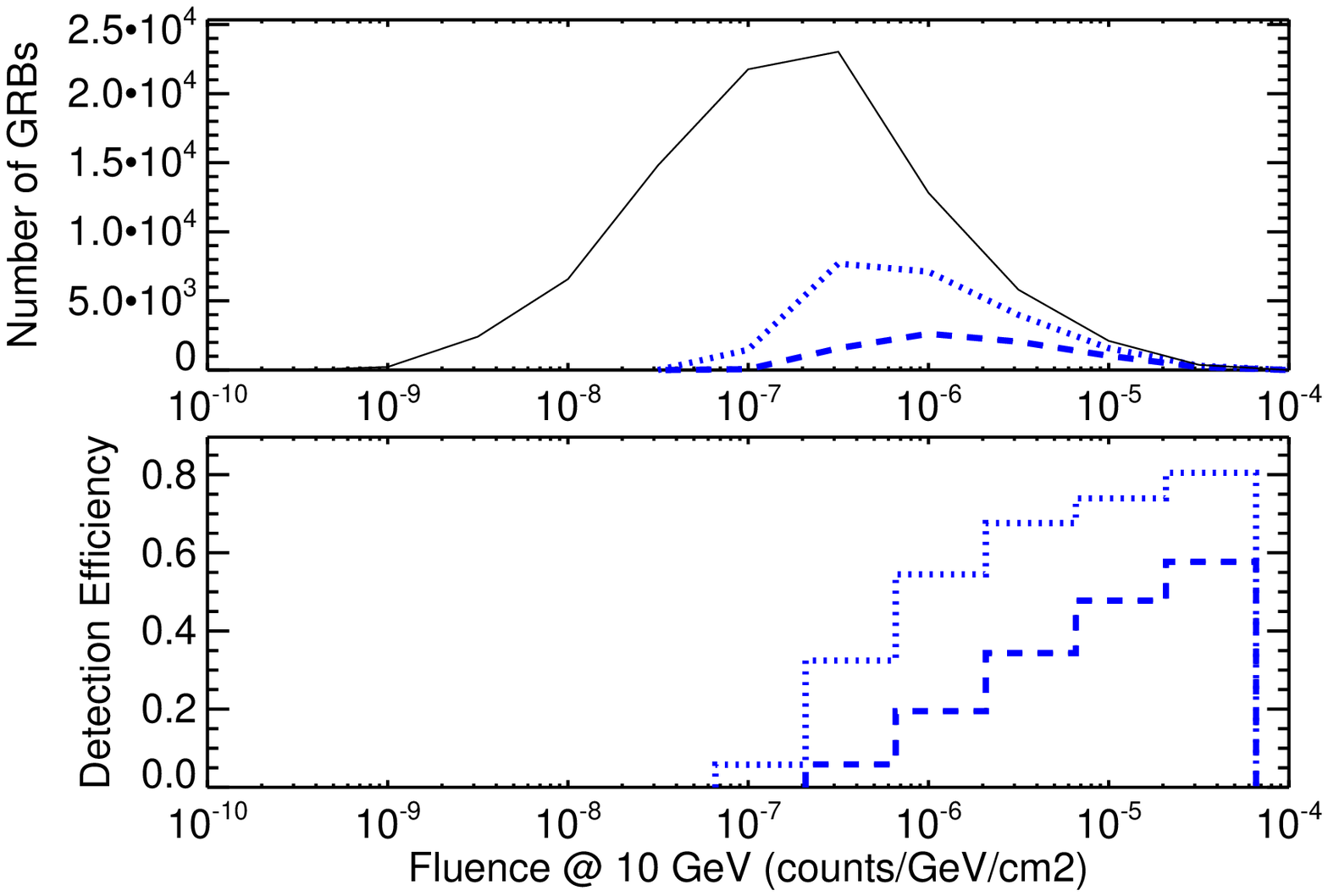}
\caption{High energy fluence distribution of GRBs in our bandex (left) and fixed (right) models, together with the distribution for detected GRBs.  Lower panels are the detected fraction of GRBs in a given bin.  Solid black and dashed blue lines are distributions for baseline effective area, while dotted lines are the corresponding values for the optimistic area function.  The thin solid lines in the top panels are the distribution for the full population.}
\label{fig:ctaflu}
\end{figure*}

In this Appendix, we examine in more detail how the population of GRBs that pass our detection criteria compares to the entire population of simulated {\it Swift}-like GRBs.  This will give us some insight as to the properties that might be expected of a burst with a confident CTA detection.  It will also be useful to look at how the assumption of different effective area functions can affect results.

The distribution of integration timescales that maximize detection significance is shown in Figure \ref{fig:timescales}.  The two spectral models produce similar results in this distribution.  This result suggests that a integration timescale of 100 to 500 seconds after the commencement of ground-based observation will be favored for GRB detection in most cases, assuming a universal $t^{-1.5}$ falloff in the afterglow lightcurve.  A small subset of bright GRBs however are still visible against the background some hours after the event trigger ($10^4$ seconds in the longest timescale considered here), and nonzero results are found for all bins in Fig.~\ref{fig:timescales}.  On the right hand side of this plot, we show how detection efficiency varies with GRB T90 duration, as determined by BATSE.  Not surprisingly, longer bursts always have a better chance of being detected, but the majority of detected GRBs have T90 values from 30 to 100 sec, due to the scarcity of bursts with T90 $>$ 100 sec.

\begin{figure}
\psfig{file=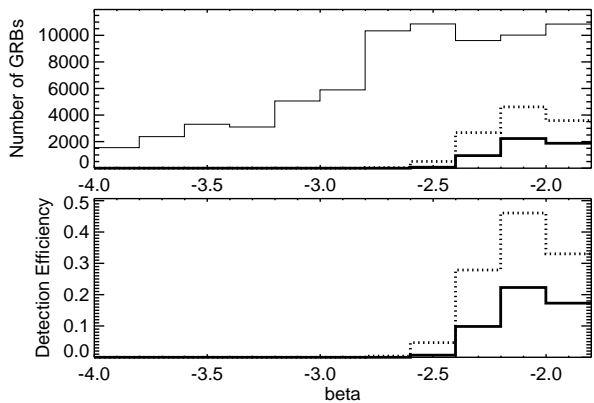,width=\figwidth}
\caption{The distribution of beta parameters from the bandex model.  As in previous figures, the thin line is for the whole population, the thick solid line shows the GRBs detected using a baseline effective area, and the dotted line shows detections using an optimistic effective area function.  The lower panel shows the fraction of detected GRBs in each bin.  The rightmost bin in each panel designates GRBs that had $\beta > -2$ in the BATSE sample, and have been reset to -2 for this calculation.}
\label{fig:ctabeta}
\end{figure}

The differences between the fixed and bandex models become most apparent when we consider the distribution in high energy fluences predicted by each, as are shown in Figure \ref{fig:ctaflu}.  In general, the bandex model has a much wider distribution in high energy fluence, because the beta parameter introduces another degree of freedom into the extrapolation, and steep beta indices lead to a subset of the bursts in the sample having extremely low levels of high energy emission.  Conversely, the brightest bandex GRBs are brighter than the brightest bursts in the fixed model, as the latter are limited to a fluence ratio of 0.1 between $\sim$1 MeV and $\sim$1 GeV, while the corresponding ratio in the bandex model can be as high as 1.0, with $\beta =-2$.  This accounts for our somewhat unexpected result that while overall detection rates predicted by the bandex model are lower, detected GRBs in this model tend to be brighter than for the fixed model.  Figure \ref{fig:ctabeta} shows the distribution of $\beta$ indices for detected GRBs in the bandex model.  Only GRBs with fairly hard extrapolated spectra, $\beta \gtrsim -2.5$, are capable of being detected.

\begin{figure}
\psfig{file=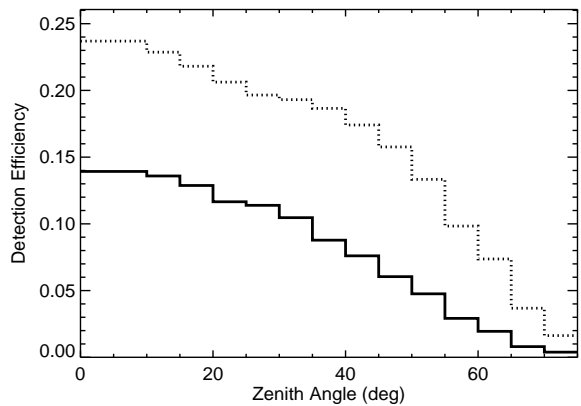,width=\figwidth}
\caption{The fraction of GRBs detected as a function of zenith angle for the bandex model with baseline (solid) and optimistic (dotted) effective area functions.  Results for the fixed model are qualitatively similar.}
\label{fig:ctatheta}
\end{figure}

Figure \ref{fig:ctatheta} shows how the probability of detecting a GRB varies with the zenith angle $\theta_{zen}$ at which it is observed.  GRBs in our model are assumed to be observed at a single instantaneous point relative to zenith.  While motion on the sky over the observation period $T_{obs}$ will change $\theta_{zen}$ over the course of longer integrations, the effect is small enough that we ignore it here.  Detection efficiency is a weak function of $\theta_{zen}$ out to $\sim$40 degrees, where increasing energy threshold is compensated by increasing solid angles.  At higher angles the detection efficiency declines more quickly.  However, GRBs can in principle be detected out to angles as large as 70 degrees, where the energy threshold is raised by a factor of 25 (Eq.~\ref{eq:zenshift}).  These would have to be at low redshift, so as not to be completely obscured by EBL opacity combined with the elevated energy threshold of the telescope.

\section[]{Prompt phase observations}
\label{sec:promptonly}

\begin{figure}
\psfig{file=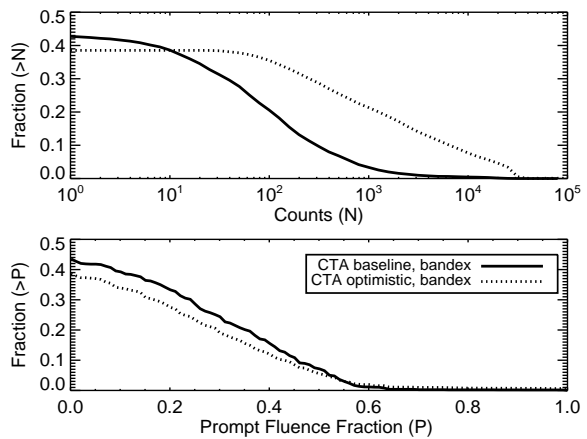,width=\figwidth}
\caption{{\bf Top:}  The integral distribution of photon counts arising from the prompt phase for detected GRBs in our bandex model.  The solid line is for a baseline effective area function, while the dotted line is for the optimistic.  Note that the y-axis intercept indicates the fraction of GRBs for which any photons are detected during the prompt phase; the majority of GRBs are detected purely on the basis of afterglow fluence.  Results with the fixed model are qualitatively similar.  {\bf Bottom:} The integral distribution of the fraction of high-energy fluence collected during the T90 period for detected GRBs; the remainder of the fluence being due to the burst afterglow. }
\label{fig:ctaprompt}
\end{figure}

\begin{figure*}
\plottwo{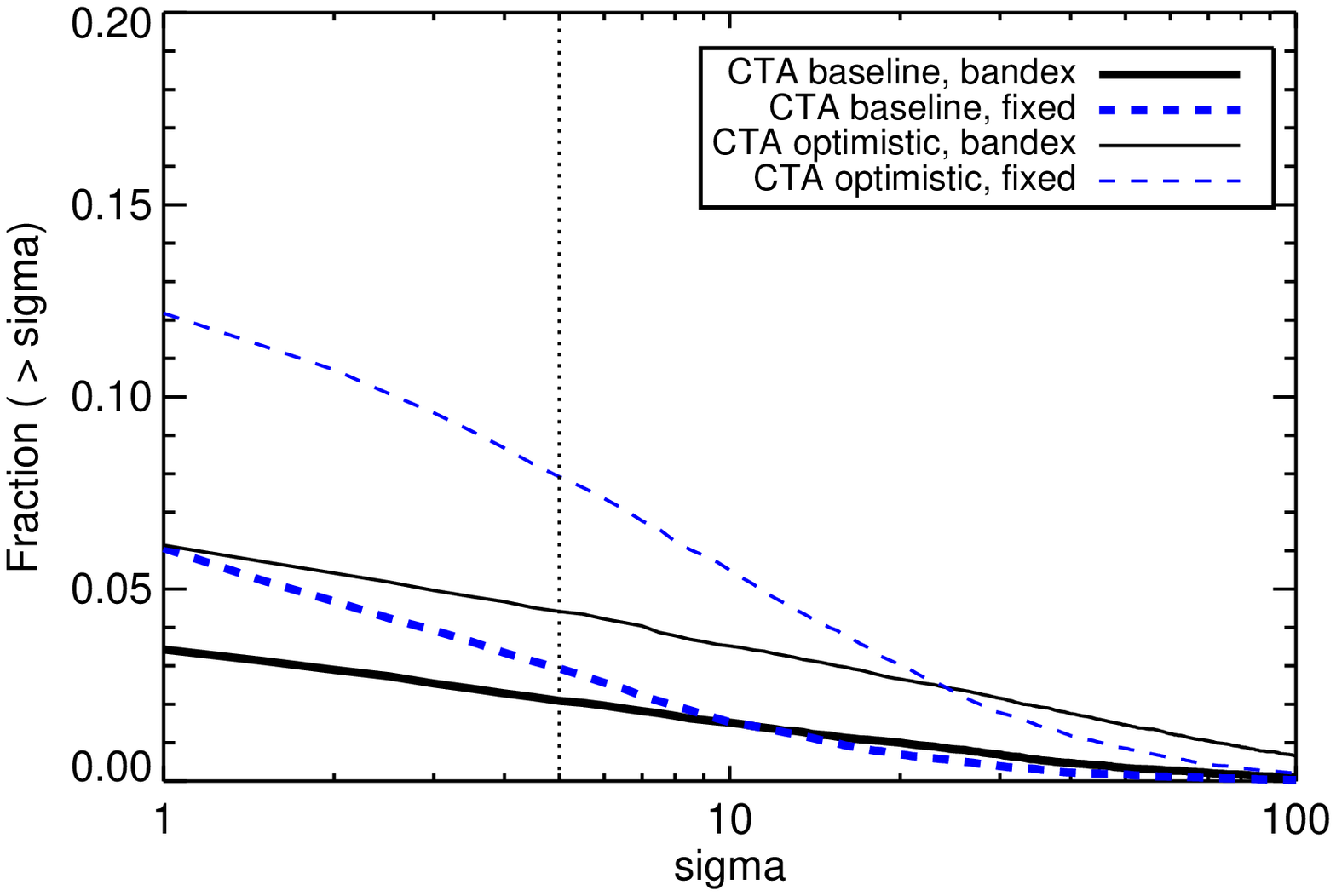}{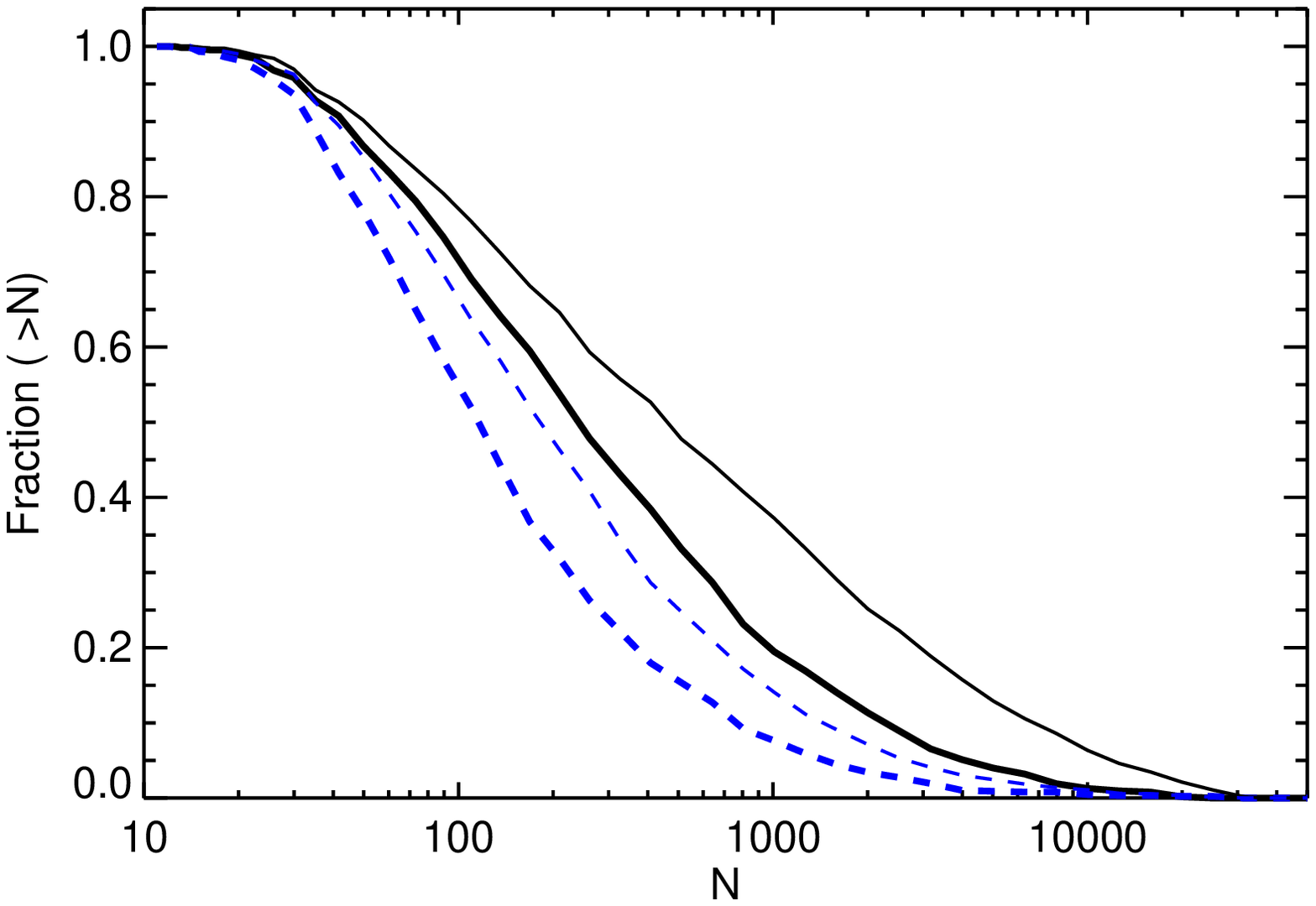}
\caption{The integral distributions of detection significances (left) and photon counts for detected GRBs (right), in a scenario in which GRBs only emit during the prompt (T90) phase.  Line types are as in Fig. \ref{fig:ctasigma}.
}
\label{fig:ctapromptgamcnt}
\end{figure*}

GRB detection in our calculation is heavily reliant on emission during the early afterglow phase.  Only about 21 percent of GRBs in our sample have prompt emission (T90) phases longer than 60 sec, which we assume as a typical delay time for observations with the LSTs.  The majority of GRBs are therefore completely inaccessible during the prompt phase for the standard assumption of a 60 sec time delay.    
As shown in Figure \ref{fig:timescales}, there is a definite bias toward longer duration GRBs in the detected portion of the population.  Figure \ref{fig:ctaprompt} summarizes the amount of fluence in detected GRBs that arises from $t<\mbox{T90}$.  About 57 percent of bursts detected with a baseline effective area have no prompt phase fluence, while only about 10 percent have more than half the detected fluence arising from emission during the prompt period.  With an optimistic effective area function, the fraction of GRBs seen purely in the afterglow period is slightly higher.

We can also consider an extreme possibility in our detection efficiency calculation: one in which no high energy emission emerges after the prompt phase, or equivalently, the light curve index $\gamma$ in Eq. \ref{eq:lci} is taken to $+\infty$.  This is found to reduce detection efficiencies to about one-third their values in the standard calculation: 0.027 and 0.040 for the bandex and fixed model with baseline effective area (0.057 and 0.11 in the optimistic case).  Fig. \ref{fig:ctapromptgamcnt} shows the distribution of sigma values and counts for detected GRBs in such a case.  These can be compared to those predicted in Fig. \ref{fig:ctasigma}.

\end{document}